\documentclass[a4paper,11pt]{article}
\usepackage[english]{babel}
\pdfoutput=1 

\usepackage{jheppub} 

\usepackage[T1]{fontenc} 

\usepackage[T1,T2A]{fontenc}
\usepackage[utf8x]{inputenc}

\usepackage{ytableau}
\ytableausetup{smalltableaux}
\RequirePackage[vcentermath]{youngtab}

\newcommand{\bi}{\begin{itemize}}
    \newcommand{\ei}{\end{itemize}}
\newcommand{\bea}{\begin{eqnarray}}
    \newcommand{\eea}{\end{eqnarray}}
\newcommand{\bt}{\begin{tabular}}
    \newcommand{\et}{\end{tabular}}
\newcommand{\bc}{\begin{center}}
    \newcommand{\ec}{\end{center}}

\newcommand{\be}{\begin{equation}}
    \newcommand{\ee}{\end{equation}}
\newcommand{\ba}{\begin{array}}
    \newcommand{\ea}{\end{array}}

\def\bbox{{\,\lower0.9pt\vbox{\hrule \hbox{\vrule height 0.2 cm
                \hskip 0.2 cm \vrule height 0.2 cm}\hrule}\,}}
\newcommand{\dsl}{\pa \kern-0.5em /}

\usepackage{etoolbox}

\makeatletter

\def\env@sqcases{%
    \let\@ifnextchar\new@ifnextchar
    \left\lbrack
    \def\arraystretch{1.2}%
    \array{@{}l@{\quad}l@{}}%
}
\makeatother

\usepackage{tikz}

\makeatletter \@addtoreset{equation}{section} \makeatother

\def\slashchar#1{\setbox0=\hbox{$#1$}           
    \dimen0=\wd0                                 
    \setbox1=\hbox{/} \dimen1=\wd1               
    \ifdim\dimen0>\dimen1                        
    \rlap{\hbox to \dimen0{\hfil/\hfil}}      
    #1                                        
    \else                                        
    \rlap{\hbox to \dimen1{\hfil$#1$\hfil}}   
    /                                         
    \fi}

\pdfoutput=1

\title{\boldmath Off-shell invariants of linearized $4D, \mathcal{N}=2$  supergravity in the harmonic approach}

\author[a,b]{Evgeny~Ivanov,}
\author[a,b]{Nikita~Zaigraev}

\affiliation[a]{Bogoliubov Laboratory of Theoretical Physics, JINR,\\141980 Dubna, Moscow region, Russia}
\affiliation[b]{Moscow Institute of Physics and Technology,\\ 141700 Dolgoprudny, Moscow region, Russia}

\emailAdd{eivanov@theor.jinr.ru}
\emailAdd{nikita.zaigraev@phystech.edu}

\abstract{Using the harmonic superspace approach, we construct, at the linearized level, $\mathcal{N}=2$ supersymmetric curvatures generalizing scalar curvature, Ricci curvature and Weyl tensor. These supercurvatures are the
building blocks of various linearized $4D, \, \mathcal{N}=2$ Einstein supergravity invariants. The supercurvatures involving the scalar and Ricci curvatures are analytic harmonic ${\cal N}=2$ superfields,
 while the Weyl supertensor is a chiral $\mathcal{N}=2$ superfield. As the basic distinguished feature of our construction, all these objects are
expressed through the fundamental analytic gauge prepotentials $h^{++M}, M= (\alpha\dot\alpha, \alpha+, \dot\alpha+, 5).$
The related characteristic  features are the heavy use of harmonic derivatives and harmonic zero-curvature equations.
On a number of instructive examples, we describe the component reduction of the  superfield invariants constructed. }

\makeatletter
\gdef\@fpheader{}
\makeatother

\begin{document}
\maketitle
\flushbottom

\section{Introduction}

Among various generalizations of Einstein's theory of gravity, supergravity and higher-curvature gravity are of most attention for long. Supergravity naturally unifies fields of different spins within a single theory  and
leads to essential improvements of UV  behavior (see, e.g., \cite{Freedman:2012zz, Deser:1977yyz}). Theories with higher curvatures also exhibit a number of non-trivial quantum properties, e.g.: (a) modification of the
Gilbert-Einstein action by terms quadratic in curvature yields a renormalizable theory \cite{Stelle:1976gc}; (b) the $R+R^2$ Starobinsky theory \cite{Starobinsky:1980te} is ghost free \cite{Whitt:1984pd} and is exploited in
models of cosmological inflation; (c) the pure $R^2$ gravity (without $R$-term) is also ghost-free \cite{Alvarez-Gaume:2015rwa}. It is of significant interest to study the supergravity theories involving  higher-curvature
terms in the bosonic limit \cite{Cecotti:1987sa}. The invariants of the  higher-derivative supergravities find applications as possible supergravity counterterms, are relevant to the black hole entropy problem, have
applications in cosmology and also provide information on the relevant conformal anomalies. Supersymmetric generalizations of the terms constructed from curvatures are also of interest on their own as they can shed more
light on the intrinsic algebraic and geometric structure of diverse off-shell formulations  of supergravity. An overview of applications of the  higher-order supergravity invariants and a detailed discussion of methods of
their construction can be found, e.g., in ref. \cite{Ozkan:2024euj}.

The most efficient and convenient way to deal with supersymmetric theories is  the superspace approach \cite{Wess:1992cp, 1001, Buchbinder:1998qv}.
In this paper, we consider linearized higher-order invariants of $\mathcal{N}=2$ supergravity in  the harmonic superspace \cite{Galperin:1984av}.
Among all the covariant superspace approaches to $\mathcal{N}=2$ supergravity (for review, see \cite{Kuzenko:2022ajd}) the formulation in harmonic superspace has a number of advantages.
The main merit of such a formulation is the manifest presence of $\mathcal{N}=2$ supergravity prepotentials, which have a natural geometric interpretation
as vielbeins covariantizing the derivatives with respect to harmonic variables. This way, the harmonic formulation of  $\mathcal{N}=2$ supergravities
non-trivially generalizes the geometric Ogievetsky-Sokatchev approach to $\mathcal{N}=1$ supergravity \cite{Ogievetsky:1978mt}.

The $\mathcal{N}=2$ supergravity \textit{unconstrained prepotentials} appear as analytic vielbeins in the covariant harmonic derivative which preserves  the notion of harmonic analyticity:
\begin{equation}
\mathfrak{D}^{++} = \mathcal{D}^{++} + h^{++\alpha\dot{\alpha}} \partial_{\alpha\dot{\alpha}}
+
h^{++\alpha+}\partial^-_\alpha
+
h^{++\dot{\alpha}+}\partial^-_{\dot{\alpha}}
+
h^{++5}\partial_5.
\end{equation}
The prepotentials with a clear geometric origin remain ``terra incognita'' in all other formulations of $\mathcal{N}=2$ supergravity\footnote{The basic difference from the Mezincescu-type \cite{Mez} prepotentials is
that the latter carry some gauge invariances of non-geometric character which admit no a direct generalization to the full nonlinear level (see, e.g., discussion in \cite{18}).}. Meanwhile, the presence of prepotentials is a
necessary ingredient for studying the equations of motion and quantizing the theory in a manifestly supersymmetric manner. Another striking feature of the harmonic superspace is the possibility to construct actions that do
not admit formulations in the conventional superspaces.

Despite the active developments of superfield formulations of extended supergravity theories for recent years, not too many studies were specially devoted
to the harmonic superspace formulation of $\mathcal{N}=2$ supergravity and so many relevant questions remained opened. In this regard,
it makes sense to start by briefly listing the main works on the subject (for a comprehensive review, see \cite{Ivanov:2022vwc, 18}).

$\mathcal{N}=2$ harmonic superspace was introduced in \cite{Galperin:1984av} as the most adequate superspace for description of general theories with manifest off-shell $\mathcal{N}=2$ supersymmetry. In
\cite{Galperin:1984av}, $\mathcal{N}=2$ Einstein  supergravity analytic prepotentials and the corresponding gauge group were discovered. The invariant nonlinear action for $\mathcal{N}=2$ Einstein supergravity was built by
Galperin, Ky and Sokatchev in \cite{Galperin:1987em}. In the accompanying work \cite{Galperin:1987ek} there was constructed a multiplet of $\mathcal{N}=2$ conformal supergravity ($\mathcal{N}=2$ Weyl multiplet) in harmonic
superspace and various versions of $\mathcal{N}=2$ Einstein supergravity were recovered through the appropriate superconformal harmonic compensators. In particular, the ``principal'' version of $\mathcal{N}=2$ Einstein
supergravity with an infinite number of auxiliary fields was introduced. In this version, an off-shell hypermultiplet is used as a superconformal compensator (in addition to the $\mathcal{N}=2$ vector multiplet) and just
this version permits the most general interaction Lagrangians of hypermultiplet matter in the $\mathcal{N}=2$ supergravity background. The harmonic superspace construction of general off-shell hypermultiplet couplings in $\mathcal{N}=2$
supergravity background and their relation to quaternionic-K\"{a}hler geometry were elaborated in a series of papers \cite{Galperin:1992pj, Ivanov:1998de, Ivanov:1998ih, Ivanov:1999vg}. The linearized action for
$\mathcal{N}=2$ Einstein supergravity was constructed by Zupnik \cite{Zupnik:1998td} and recently it was generalized to $\mathcal{N}=2$ higher spin supergravities in \cite{Buchbinder:2021ite}. The interrelations between
$\mathcal{N}=2$ supercurrents and supergravity prepotentials were discussed in \cite{Kuzenko:1999pi, Butter:2010sc}. More recently \cite{Butter:2015nza}, there was suggested a new covariant approach to the harmonic
superspace as a symbiosis  of the latter with conformal $4D, {\cal N}=2$ superspace \cite{Butter:2011sr}. Nevertheless, many issues related to the harmonic superspace formulation of $\mathcal{N}=2$ supergravity have never been addressed. In particular, this concerns the construction of off-shell higher-derivative
$\mathcal{N}=2$ supergravity invariants in the harmonic approach.

On the other hand, quite a few remarkable results in this area were achieved in other superfield approaches to $\mathcal{N}=2$ supergravity.
The $\mathcal{N}=2$ curvature-squared  \textit{off-shell} invariants   were constructed in \cite{Butter:2013lta} and \cite{Kuzenko:2015jxa}
(see also review \cite{Kuzenko:2017uqg}), based upon the formulation
of $\mathcal{N}=2$ conformal supergravity in $SU(2)$ superspace \cite{Grimm:1980kn, Kuzenko:2008ep} and $\mathcal{N}=2$ conformal superspace \cite{Butter:2011sr}.
The $\mathcal{N}=2$ supersymmetrizations of $R^4$ invariants were considered in \cite{Moura:2002ip, deWit:2010za}.
Using \textit{the logarithm construction}, proposed in  \cite{Butter:2013lta}  (for a review see  section 10.2 of \cite{Kuzenko:2022ajd})
one can set up some higher-order invariants. For example, higher-derivative super-Weyl $\mathcal{N}=2$ invariants, containing $F^{2n}$, with $n=3,4,\dots$
were proposed in  \cite{Kuzenko:2013gva}. Another mechanism of generating higher-derivative couplings in $\mathcal{N}=2$ supergravity is \textit{the procedure
of constructing composite reduced chiral superfields} \cite{Butter:2010jm} (for a brief review see section 8.4 of \cite{Kuzenko:2022ajd}).
However, to the best of our knowledge, specific explicit examples of higher-derivative $\mathcal{N}=2$ supergravity invariants beyond $R^2$
were not so far constructed  in this approach.

\medskip

Note that the construction of the {\it on-shell} $\mathcal{N}$-extended supergravity invariants in  harmonic superspace have also been extensively discussed
in a number of papers devoted to the on-shell extended supergravity counterterms. For example, in \cite{Drummond:2003ex} some invariants of $4D, \mathcal{N}=8$ linearized supergravity were constructed;
the on-shell harmonic superspace linearized counterterms ($\partial^{2(\mathcal{N}-4)} R^4$) for $\mathcal{N} \geq 4$ supergravities were presented in \cite{Bossard:2011tq}.
A short summary of the results on such invariants can be found in a review \cite{Ozkan:2024euj}.  For a recent discussion of these on-shell superinvariants
at the nonlinear level, see, e.g., \cite{Kallosh:2023pkr}.

\medskip

In this paper, we construct off-shell higher-derivative $\mathcal{N}=2$ supergravity invariants in harmonic superspace at the linearized level, consistently applying the whole
machinery of the harmonic approach. The main distinction of this scheme from the more conventional superspace methods
\cite{Butter:2013lta, Kuzenko:2015jxa, Kuzenko:2017uqg, Moura:2002ip, deWit:2010za}  is that its basic building blocks are directly related to the unconstrained
\textit{analytic prepotentials} underlying the structure of curved $\mathcal{N}=2$ harmonic superspace and encapsulating the off-shell supergravity
multiplet in Wess-Zumino (WZ) gauge. This radically simplifies deducing the component forms of the invariants. Though the most difficult part of the road to components remains
solving the so called harmonic zero-curvature equations, it proves to be rather straightforward in WZ gauge.

As we shall show, the basic objects in constructing
$\mathcal{N}=2$ higher-derivative supergravity invariants are linearized gauge-invariant supercurvatures which have a surprisingly simple structure:
\begin{equation}
\mathcal{F}^{++\alpha\dot{\alpha}} = (\mathcal{D}^+)^4 G^{--\alpha\dot{\alpha}}
=
- 8i \theta^+_\rho \bar{\theta}^+_{\dot{\rho}} \left( \mathcal{R}^{(\alpha\rho)(\dot{\alpha}\dot{\rho})} - \frac{1}{8} \epsilon^{\alpha\rho}\epsilon^{\dot{\alpha}\dot{\rho}} R
\right) + \dots, \label{1.2}
\end{equation}
\begin{equation}
\mathcal{F}^{++5} = (\mathcal{D}^+)^4 G^{--5}
= \frac{i}{2} (\theta^+)^2 R - \frac{i}{2} (\bar{\theta}^+) R  + \dots.,
\end{equation}
\begin{equation}
\mathcal{W}_{(\alpha\beta)} = (\bar{\mathcal{D}}^+)^2 \left(\mathcal{D}^+_{(\alpha} G^{---}_{\beta)} + \mathcal{D}^-_{(\alpha} G^{--+}_{\beta)}
-
\partial_{(\alpha}^{\dot{\rho}}G^{--}_{\beta)\dot{\rho}} \right) =  32 \theta^{-(\gamma} \theta^{+\delta)} \mathcal{R}_{(\alpha\beta\gamma\delta)}
+
\dots.
\end{equation}
The linearized $\mathcal{N}=2$ supercurvatures $\mathcal{F}^{++\alpha\dot{\alpha}}$ and $\mathcal{F}^{++5}$ are analytic superfields, while the linearized $\mathcal{N}=2$ Weyl tensor $\mathcal{W}_{(\alpha\beta)}$ is a chiral
superfield. All these supercurvatures are constructed out of the superfield potentials $G^{--\alpha\dot{\alpha}}, G^{--5}, G^{--\alpha\pm} $ related to the basic analytic prepotentials via the well defined harmonic
zero-curvature equations. In WZ gauge, these equations can be directly solved, giving rise to the explicit expressions for the $G$-potentials in terms of the $\mathcal{N}=2$ supergravity
multiplet fields and their derivatives.  In their component expansion, the supercurvatures involve all possible irreducible linearized gravity curvatures quadratic in derivatives:
the scalar curvature $R$, the irreducible part of Ricci tensor $\mathcal{R}_{(\alpha\beta)(\dot{\alpha}\dot{\beta})}$
and the linearized Weyl tensor $\mathcal{R}_{(\alpha\beta\gamma\delta)}$. So these supercurvatures  can be viewed  as the building blocks of various linearized $\mathcal{N}=2$ supergravity invariants.

\medskip

The paper is organized as follows. In section
\ref{sec:2} we introduce notations and shortly recall the structure of harmonic superspace. In section \ref{sec:3} we describe prepotentials
of $\mathcal{N}=2$ Einstein supergravity multiplet and the corresponding gauge group. Next we perform  the linearization procedure and study in detail  the
component contents of the analytic prepotentials. In section \ref{sec:4}, the covariant $\mathcal{N}=2$ superfields $G^{\pm\pm M}$ are introduced and the linearized $\mathcal{N}=2$ supersymmetric
Einstein action is constructed.
In section \ref{sec:5}, the linearized supercurvatures are defined and all invariant quadratic actions in harmonic superspace are presented.
In section \ref{sec:6}, we elaborate on the component reduction of the linearized $\mathcal{N}=2$ Einstein supergravity, obtain the linearized supercurvatures
and the linearized quadratic invariants. In section \ref{sec:7} we discuss construction of more general linearized $\mathcal{N}=2$ supergravity invariants.
Section \ref{sec:8} is for conclusions and outlook. In particular, we briefly sketch the higher-spin generalizations of the invariants constructed.

Three Appendices collect some technical details. Appendix \ref{sec:A1} provides the linearized gravity invariants in the spinor notations used throughout  the paper. In Appendix  \ref{sec:A2}
we introduce the linearized spin $\frac{3}{2}$ invariants which are of need, when performing the component reduction in the fermionic sector.
In Appendix \ref{app: zero curv solutions} we present the solutions of the harmonic zero-curvature equations, which, as was already mentioned, are a necessary  part of passing
to the component Lagrangians.

\section{Harmonic superspace}\label{sec:2}

We proceed from $\mathcal{N}=2$ harmonic superspace in the analytic basis, with an extra $x^5$ coordinate (for a pedagogical introduction,
see \cite{18})\footnote{Here we use standard notations: $x^{\alpha\dot{\alpha}} = x^m (\tilde{\sigma}_m)^{\alpha\dot{\alpha}}$, $\partial_{\alpha\dot{\alpha}} = \frac{1}{2}\sigma^m_{\alpha\dot{\alpha}}\partial_m$ (so $\partial_{\alpha\dot{\alpha}}x^{\beta\dot{\beta}} = \delta^\beta_\alpha \delta^{\dot{\beta}}_{\dot{\alpha}}$, $\partial_{\alpha\dot{\beta}}\partial^{\dot{\beta}\beta}=\frac{1}{4}\delta_\alpha^\beta \Box$), $\hat{\alpha} = (\alpha, \dot{\alpha})$, $\partial^-_{\hat{\alpha}} = \frac{\partial}{\partial \theta^{+\hat{\alpha}}}$,
$\partial^+_{\hat{\alpha}} = \frac{\partial}{\partial \theta^{-\hat{\alpha}}}$, $(\psi\chi)
= \psi^\alpha \chi_\alpha$, $(\bar{\psi}\bar{\chi}) = \bar{\psi}_{\dot{\alpha}} \chi^{\dot{\alpha}}$, $(\theta^{\hat{+}})^2 = \theta^{+\hat{\alpha}} \theta^+_{\hat{\alpha}}$  etc.}:
\begin{equation}\label{eq: HS coord}
\left\{x^{\alpha\dot{\alpha}}, \theta^{+\hat{\alpha}}, \theta^{-\hat{\alpha}}, u^{\pm}_i, x^5\right\}, \quad \hat{\alpha} = (\alpha, \dot\alpha)\,.
\end{equation}
The harmonics $u^\pm_i$ are auxiliary coordinates that  parametrize an internal sphere $S^2 \sim SU(2)/U(1)$ and  satisfy the relation $u^{+i}u^-_i = 1$.
The presence of harmonics $u^\pm_i$ in the coordinate set inevitably leads to an infinite number of component fields in any unconstrained harmonic superfield.
The additional coordinate  $x^5$ is needed to introduce the interaction of $\mathcal{N}=2$ Maxwell multiplet with hypermultiplet, as well as to describe massive hypermultiplet
(with a mass equal to central charge). In this paper, all  harmonic superfields are assumed to be $x^5$-independent. The coordinate $x^5$ will be merely used  for the
purpose to simplify handling $\mathcal{N}=2$ supergravity prepotentials in the linearized theory.
In the analytic basis, rigid $\mathcal{N}=2$ supersymmetry transformations act on the coordinates \eqref{eq: HS coord} as:
\begin{equation}\label{eq: SUSY transf}
\delta_\epsilon x^{\alpha\dot{\alpha}}
=
-4i \left(\epsilon^{-\alpha} \bar{\theta}^{+\dot{\alpha}} + \theta^{+\alpha} \bar{\epsilon}^{-\dot{\alpha}} \right),
\quad
\delta_\epsilon \theta^{\pm\hat{\alpha}} = \epsilon^{\pm\hat{\alpha}},
\quad
\delta_\epsilon u^{\pm}_i = 0,
\quad
\delta_\epsilon x^5 =
2i \epsilon^{-\hat{\alpha}}\theta^+_{\hat{\alpha}}.
\end{equation}
Here $\epsilon^{\pm\hat{\alpha}} := \epsilon^{\hat{\alpha}i}u^\pm_i$ are relevant spinorial parameters.
These transformations preserve an analytic subspace of harmonic superspace:
\begin{equation}\label{eq: AHSS}
\zeta: = \left\{x^{\alpha\dot{\alpha}}, \theta^{+\hat{\alpha}}, u^\pm_i\right\}.
\end{equation}
This invariant subspace contains just half the original Grassmann variables.

An analogue of complex conjugation in harmonic superspace  is the \textit{tilde conjugation}. On complex functions, it acts like the ordinary complex conjugation.
Its action on the harmonic superspace coordinates is as follows
\begin{equation}
\widetilde{x^{\alpha\dot{\alpha}}} = x^{\alpha\dot{\alpha}},
\quad
\widetilde{\theta^\pm_\alpha} = \bar{\theta}^\pm_{\dot{\alpha}},
\quad
\widetilde{\bar{\theta}^\pm_{\dot{\alpha}}} = - \theta^\pm_\alpha,
\quad
\widetilde{u^{\pm i}} = - u_i^\pm,
\quad
\widetilde{u^\pm_i} = u^{\pm i},
\quad
\widetilde{x^5} = x^5.
\end{equation}

The spinor derivatives which are covariant with respect to $\mathcal{N}=2$ supersymmetry \eqref{eq: SUSY transf} are defined as:
\begin{equation}\label{eq: spinor cov}
\begin{split}
	&\mathcal{D}^+_{\hat{\alpha}} := \partial^+_{\hat{\alpha}},
	\quad
	\\&
	\mathcal{D}^-_\alpha
	:=
	-\partial^-_\alpha + 4i \bar{\theta}^{-\dot{\alpha}} \partial_{\alpha\dot{\alpha}}
	-
	2i \theta^-_\alpha \partial_5,
	\quad
	\\
	&
	\bar{\mathcal{D}}^-_{\dot{\alpha}} := - \partial^-_{\dot{\alpha}} - 4i \theta^{-\alpha}
	\partial_{\alpha\dot{\alpha}} +2i \bar{\theta}^-_{\dot{\alpha}} \partial_5.
\end{split}
\end{equation}
The covariant harmonic derivatives are given by:
\begin{equation}\label{eq:D0}
\begin{split}
	&\mathcal{D}^{++} = \partial^{++} - 4i \theta^{+\rho} \bar{\theta}^{+\dot{\rho}} \partial_{\rho\dot{\rho}} + \theta^{+\hat{\rho}} \partial^+_{\hat{\rho}}
	+ i\theta^{+\hat{\rho}}\theta^+_{\hat{\rho}}  \partial_5,
	\\
	&\mathcal{D}^{--} = \partial^{--} - 4i \theta^{-\rho} \bar{\theta}^{-\dot{\rho}} \partial_{\rho\dot{\rho}} + \theta^{-\hat{\rho}} \partial^-_{\hat{\rho}}
	+ i \theta^{-\hat{\rho}}\theta^-_{\hat{\rho}}  \partial_5,
	\\
	&\mathcal{D}^0 = \partial^0 + \theta^{+\hat{\rho}} \partial^-_{\hat{\rho}}
	-
	\theta^{-\hat{\rho}} \partial^+_{\hat{\rho}}.
\end{split}
\end{equation}
Here we used the following notations for the partial harmonic derivatives:
\begin{equation}
\partial^{++} = u^{+i} \frac{\partial}{\partial u^{-i}},
\qquad
\partial^{--} = u^{-i} \frac{\partial}{\partial u^{+i}},
\qquad
\partial^0 =  u^{+i} \frac{\partial}{\partial u^{+i}}
-
u^{-i} \frac{\partial}{\partial u^{-i}}.
\end{equation}
They satisfy the relations:
\begin{equation}
[   \partial^{++},  \partial^{--} ] =   \partial^{0},
\qquad
[\partial^0, \partial^{\pm\pm}] = \pm 2 \partial^{\pm\pm}.\label{PartHarm}
\end{equation}

The covariant harmonic derivatives play an important role in the formulation of $\mathcal{N}=2$ supergravity in harmonic superspace.
They satisfy the $SU(2)$-algebra relations similar to \eqref{PartHarm}:
\begin{equation}
[\mathcal{D}^{++} , \mathcal{D}^{--} ] = \mathcal{D}^0,
\qquad [\mathcal{D}^0, \mathcal{D}^{\pm\pm}] = \pm 2 \mathcal{D}^{\pm\pm}.
\end{equation}
The spinor covariant derivatives \eqref{eq: spinor cov} obey the relations:
\begin{subequations}
\begin{equation}
	[\mathcal{D}^{--}, \mathcal{D}^+_{\hat{\alpha}}] = \mathcal{D}^-_{\hat{\alpha}},
	\qquad
	[\mathcal{D}^{++}, \mathcal{D}^-_{\hat{\alpha}}] = \mathcal{D}^+_{\hat{\alpha}},
	\qquad
	[\mathcal{D}^{++}, \mathcal{D}^+_{\hat{\alpha}}] = 0,
	\qquad
	[\mathcal{D}^{--}, \mathcal{D}^-_{\hat{\alpha}}] = 0,
\end{equation}
\begin{equation}
	\{\mathcal{D}^+_\alpha, \bar{\mathcal{D}}^-_{\dot{\alpha}} \} = -4i \partial_{\alpha\dot{\alpha}} ,
	\;\;
	\{\bar{\mathcal{D}}^+_{\dot{\alpha}}, \mathcal{D}^-_\alpha \} = 4i \partial_{\alpha\dot{\alpha}},
	\;\;
	\{\mathcal{D}^+_\alpha, \mathcal{D}^-_\beta \} = 2i \epsilon_{\alpha\beta} \partial_5 ,
	\;\;
	\{\bar{\mathcal{D}}^+_{\dot{\alpha}}, \bar{\mathcal{D}}^-_{\dot{\beta}} \} =
	-2i \epsilon_{\dot{\alpha}\dot{\beta}}\partial_5.
\end{equation}
\end{subequations}

Harmonic superspace can be looked upon as the most general $4D, \mathcal{N}=2$ superspace \cite{Kuzenko:1998xm, Jain:2009aj},
and it is indispensable for formulation and study of classical and quantum $\mathcal{N}=2$ theories \cite{18}.
The analytic superspace \eqref{eq: AHSS} plays the fundamental role as it underlies the geometric structure of all $\mathcal{N}=2$ theories
(see, e.g., \cite{Buchbinder:2022kzl} for application to the construction of the hypermultiplet couplings to higher spins).
In this paper, $\mathcal{N}=2$ harmonic analyticity will be shown to be a necessary ingredient also in
constructing higher-derivative ${\cal N}=2$ supergravity invariants. In the next section, as an instructive example, we shall
expound how the analytic unconstrained prepotentials of $\mathcal{N}=2$ Einstein supergravity naturally appear in the harmonic superspace framework.
After that, we shall proceed to the analysis of the linearized theory.

\section{$\mathcal{N}=2$ Einstein supergravity multiplet in harmonic superspace}
\label{sec:3}

This section is devoted to a detailed discussion of $\mathcal{N}=2$ Einstein supergravity multiplet and gauge freedom in both the full-fledged nonlinear case and its linearized approximation.
We will also define the auxiliary fields redefinitions needed for performing the component reduction in section \ref{sec:6}.

\subsection{$\mathcal{N}=2$ supergravity multiplet: nonlinear set-up}

The gauge group of $\mathcal{N}=2$ Einstein supergravity is constituted by the analyticity-preserving transformations of harmonic superspace \cite{Galperin:1987ek, Galperin:1987em, 18}
\footnote{They can be treated as an analytic gauging of rigid $\mathcal{N}=2$ supersymmetry transformations \eqref{eq: SUSY transf}.}
\begin{equation}\label{eq: supertranslations}
\delta_\lambda x^{\alpha\dot{\alpha}} = \lambda^{\alpha\dot{\alpha}}(\zeta),
\quad
\delta_\lambda \theta^{+\hat{\alpha}} = \lambda^{+\hat{\alpha}}(\zeta),
\quad
\delta_\lambda \theta^{-\hat{\alpha}} = \lambda^{-\hat{\alpha}}(\zeta, \theta^-),
\quad
\delta_\lambda u^\pm_i = 0,
\quad
\delta_\lambda x^5 = \lambda^5(\zeta).
\end{equation}
Note that the only parameters which are not analytic are  $\lambda^{-\hat{\alpha}}$. Another peculiarity is that the harmonic variables
retain inert under gauge transformations \footnote{It is the gauge group of $\mathcal{N}=2$ conformal supergravity that affects harmonics, see \cite{Galperin:1987ek, 18, Ivanov:2022vwc}. }.
It is useful to introduce a multi-index $M:= (\alpha\dot{\alpha}, +\hat{\alpha}, -\hat{\alpha},5)$ and
denote the harmonic superspace coordinates as $z^M := (x^{\alpha\dot{\alpha}}, \theta^{+\hat{\alpha}}, \theta^{-\hat{\alpha}}, x^5 )$. Then super-diffeomorphism transformations
can be represented as the action of the differential operator:
\begin{equation}
\delta_\lambda z^M = [\hat{\Lambda}, z^M ],
\qquad
\hat{\Lambda}:= \lambda^M \partial_M.
\end{equation}
This is a very convenient notation, because, for instance, transformations of the superspace derivatives
$\partial_M := (\partial_{\alpha\dot{\alpha}}, \partial^-_{\hat{\alpha}}, \partial^+_{\hat{\alpha}}, \partial_5)$  can be succinctly written as
\begin{equation}
\delta_\lambda \partial_M = [\hat{\Lambda}, \partial_M].
\end{equation}
In particular, it implies that the harmonic derivative $\mathcal{D}^{++}$
ceases to be covariant\footnote{The rigid $\mathcal{N}=2$ supersymmetry transformations can be defined as a subgroup of \eqref{eq: supertranslations} distinguished by the
condition $\delta_\epsilon \mathcal{D}^{++}=0$.} under  supertranslations \eqref{eq: supertranslations}:
\begin{equation}
\begin{split}
	\delta_\lambda \mathcal{D}^{++} = [\hat{\Lambda}, \mathcal{D}^{++}]
	=
	&- \mathcal{D}^{++}\lambda^M \partial_M
	\\
	&
	- 4i \lambda^{+\rho} \bar{\theta}^{+\dot{\rho}} \partial_{\rho\dot{\rho}}
	- 4i \theta^{+\rho} \bar{\lambda}^{+\dot{\rho}} \partial_{\rho\dot{\rho}} + \lambda^{+\hat{\rho}} \partial^+_{\hat{\rho}}
	+ 2(\lambda^{\hat{+}}\theta^{\hat{+}}) \partial_5.
\end{split}
\end{equation}
The unwanted terms can be canceled by introducing a set of gauge prepotentials:
\begin{equation}
\mathcal{D}^{++} \quad \to \quad \mathfrak{D}^{++}
=
\mathcal{D}^{++} + h^{++\alpha\dot{\alpha}}\partial_{\alpha\dot{\alpha}} + h^{++\hat{\alpha}+}\partial^-_{\hat{\alpha}}
+
h^{++\hat{\alpha}-} \partial^+_{\hat{\alpha}}
+ h^{++5}\partial_5
\end{equation}
with the gauge freedom:
\begin{equation}\label{eq: perp GF}
\begin{cases}
	\delta_\lambda h^{++\alpha\dot{\alpha}}
	=
	\mathfrak{D}^{++} \lambda^{\alpha\dot{\alpha}}
	+4i \lambda^{+\alpha} \bar{\theta}^{+\dot{\alpha}}
	+ 4i \theta^{+\alpha} \bar{\lambda}^{+\dot{\alpha}} - \hat{\Lambda} h^{++\alpha\dot{\alpha}},
	\\
	\delta_\lambda h^{++\hat{\alpha}+} = \mathfrak{D}^{++} \lambda^{+\hat{\alpha}}
	-
	\hat{\Lambda} h^{++\hat{\alpha}+},
	\\
	\delta_\lambda h^{++\hat{\alpha}-} = \mathfrak{D}^{++}\lambda^{-\hat{\alpha}}
	-
	\lambda^{+\hat{\alpha}}     -
	\hat{\Lambda} h^{++\hat{\alpha}-},
	\\
	\delta_\lambda h^{++5} = \mathfrak{D}^{++}\lambda^5 - 2 \lambda^{+\hat{\alpha}} \theta^+_{\hat{\alpha}}
	-
	\hat{\Lambda} h^{++5}.
\end{cases}
\end{equation}
The modified derivative $\mathfrak{D}^{++}$ is covariant under \eqref{eq: supertranslations}, i.e. it satisfies:
\begin{equation}
\delta_\lambda \mathfrak{D}^{++} = [\hat{\Lambda}, \mathfrak{D}^{++}] + \delta_\lambda h^{++M}\partial_M = 0.
\end{equation}
Since the harmonic derivative $\mathcal{D}^{++}$ is real with respect to the tilde conjugation, it is natural to require that $\mathfrak{D}^{++}$ obeys the same property:
\begin{equation}
\widetilde{\mathfrak{D}^{++}} = \mathfrak{D}^{++}
\quad
\Rightarrow
\quad
\begin{aligned}
	&\widetilde{h^{++\alpha\dot{\alpha}}} = h^{++\alpha\dot{\alpha}},
	\\
	&\widetilde{h^{++5}} = h^{++5},
	\\
	&\widetilde{h^{++\alpha+}} = h^{++\dot{\alpha}+},
	\\
	&\widetilde{h^{++\dot{\alpha}+} }= -h^{++\alpha+}.
\end{aligned}
\end{equation}

Note that the transformation law of $h^{++\hat{\alpha}-}$ in \eqref{eq: perp GF} allows to impose the \textit{analytic gauge } $h^{++\hat{\alpha}-} = 0$, which expresses $\lambda^{-\hat{\alpha}}$
in terms of the analytic parameter $\lambda^{+\hat{\alpha}}$:
\begin{equation}
\mathfrak{D}^{++}\lambda^{-\hat{\alpha}} = \lambda^{+\hat{\alpha}}.
\end{equation}
In the analytic gauge, the differential operator $\mathfrak{D}^{++}$ becomes analytic, i.e. it satisfies the condition  $[\mathcal{D}^+_{\hat{\alpha}}, \mathfrak{D}^{++}]=0$.
Hereafter, we always stick to this gauge.

Notably, the gauge freedom \eqref{eq: perp GF} allows one to gauge away infinitely many fields from the prepotentials
and  to end up  with the off-shell $\mathcal{N}=2$ Einstein supergravity multiplet in the component expansion of $h^{++M}$.
In this WZ gauge we have\footnote{The numerical coefficients are chosen so as to have the field transformation laws as simple as possible, see \eqref{eq: field gauge transf}.}:
\begin{equation}\label{eq: WZ gauge}
\begin{split}
	&h_{WZ}^{++\alpha\dot{\alpha}} = -4i \theta^{+\beta}\bar{\theta}^{+\dot{\beta}} \Phi_{\beta\dot{\beta}}^{\alpha\dot{\alpha}}
	+
	16(\bar{\theta}^+)^2 \theta^{+\beta} \psi_\beta^{\alpha\dot{\alpha}i} u^-_i
	-
	16(\theta^+)^2 \bar{\theta}^{+\dot{\beta}}\bar{\psi}_{\dot{\beta}}^{\alpha\dot{\alpha}i}u^-_i
	+
	(\theta^+)^4 V^{\alpha\dot{\alpha}(ij)} u^-_i u^-_j
	\\
	&h_{WZ}^{++5} = -4i \theta^{+\beta}\bar{\theta}^{+\dot{\beta}} C_{\beta\dot{\beta}}
	+
	8(\bar{\theta}^+)^2 \theta^{+\beta} \rho_\beta^{i} u^-_i
	-
	8(\theta^+)^2 \bar{\theta}^{+\dot{\beta}}\bar{\rho}_{\dot{\beta}}^{i}u^-_i
	+
	(\theta^+)^4 S^{(ij)} u^-_i u^-_j
	\\
	&
	h_{WZ}^{++\alpha+} =  (\bar{\theta}^+)^2 \theta^+_{\beta} T^{(\alpha\beta)}
	+
	(\bar{\theta}^+)^2 \theta^{+\alpha}T
	+
	(\theta^+)^2 \bar{\theta}^+_{\dot{\beta}} P^{\alpha\dot{\beta}}
	+
	(\theta^+)^4 \chi^{\alpha i }u^-_i,
	\\
	& h_{WZ}^{++\dot{\alpha}+} = 
	(\theta^+)^2 \bar{\theta}^+_{\dot{\beta}} \bar{T}^{(\dot{\alpha}\dot{\beta})} + (\theta^+)^2 \bar{\theta}^{+\dot{\alpha}} \bar{T}
	-
	(\bar{\theta}^+)^2 \theta^+_\beta \bar{P}^{\beta \dot{\alpha}}
	+
	(\theta^+)^4 \bar{\chi}^{\dot{\alpha}i}u^-_i.
\end{split}
\end{equation}

The remaining fields are defined up to a residual gauge freedom. As usual, it is specified by the requirement of preservation of WZ gauge, i.e. $\delta^{res}_\lambda h^{++M} \sim h^{++M}_{WZ}$.
As a result, the set of residual parameters is obtained:
\begin{equation}\label{WZ res}
\begin{cases}
	\lambda_{WZ}^{\alpha\dot{\alpha}} = a^{\alpha\dot{\alpha}}(x)  - 4i \theta^{+\alpha} \bar{\epsilon}^{\dot{\alpha}i}(x) u^-_i - 4i \epsilon^{\alpha i}(x) u^-_i \bar{\theta}^{+\dot{\alpha}},
	\\
	\lambda_{WZ}^{+\alpha} = \epsilon^{\alpha i}(x) u^+_i
	+
	\theta^{+\beta}l_{(\beta}^{\;\;\alpha)}(x)
	+
	4i \theta^{+\beta}\bar{\theta}^{+\dot{\beta}} (\delta_{\beta}^\rho\delta_{\dot{\beta}}^{\dot{\rho}} + \Phi_{\beta\dot{\beta}}^{\rho\dot{\rho}}) \partial_{\rho\dot{\rho}} \epsilon^{\alpha i}(x) u^-_i,
	\\
	\lambda_{WZ}^{+\dot{\alpha}} = \bar{\epsilon}^{\dot{\alpha} i}(x) u^+_i
	+
	\bar{\theta}^{+\dot{\beta}}l_{(\dot{\beta}}^{\;\;\dot{\alpha})}(x)
	+
	4i \theta^{+\beta} \bar{\theta}^{+\dot{\beta}}(\delta_{\beta}^\rho\delta_{\dot{\beta}}^{\dot{\rho}} + \Phi_{\beta\dot{\beta}}^{\rho\dot{\rho}})\partial_{\rho\dot{\rho}}\bar{\epsilon}^{\dot{\alpha} i}(x) u^-_i,
	\\
	\lambda^5_{WZ} = c(x) + 2i \epsilon^{\hat{\alpha}i}(x) u^-_i \theta^+_{\hat{\alpha}}.
\end{cases}
\end{equation}
The realization of these transformations on the WZ fields directly follows from \eqref{eq: perp GF}.
As an example, let us quote  the transformations of the spin $2$, $\frac{3}{2}$, $1$ gauge fields:
\begin{subequations}
\begin{equation}
	\begin{split}
		\delta\Phi_{\beta\dot{\beta}}^{\alpha\dot{\alpha}}
		=&
		(\delta_{\beta}^\rho\delta_{\dot{\beta}}^{\dot{\rho}} + \Phi_{\beta\dot{\beta}}^{\rho\dot{\rho}}) \partial_{\rho\dot{\rho}} a^{\alpha\dot{\alpha}}
		-
		a^{\rho\dot{\rho}} \partial_{\rho\dot{\rho} } \Phi_{\beta\dot{\beta}}^{\alpha\dot{\alpha}}
		-
		\left(l_{(\beta}^{\;\;\rho)} \delta^{\dot{\alpha}}_{\dot{\rho}}
		+
		\delta^\alpha_\rho \bar{l}_{(\dot{\beta}}^{\;\;\dot{\rho})}
		\right) \left(\delta_\rho^\alpha \delta_{\dot{\rho}}^{\dot{\alpha}} + \Phi_{\rho\dot{\rho}}^{\alpha\dot{\alpha}} \right)
		\\
		&
		-2i \epsilon_{\beta i} \bar{\psi}^{\alpha\dot{\alpha}i}_{\dot{\beta}}
		-
		2i \bar{\epsilon}_{\dot{\beta}i} \psi_\beta^{\alpha\dot{\alpha}i},
	\end{split}
\end{equation}
\begin{equation}
	\begin{split}
		\delta \psi^{\alpha\dot{\alpha}i}_\beta
		=
		&-
		\epsilon^{\dot{\alpha}\dot{\beta}}
		(\delta_{\beta}^\rho\delta_{\dot{\beta}}^{\dot{\rho}} + \Phi_{\beta\dot{\beta}}^{\rho\dot{\rho}}) \partial_{\rho\dot{\rho}} \epsilon^{\alpha i}
		+
		i \frac{1}{2} T^{\;\alpha)}_{(\beta} \bar{\epsilon}^{\dot{\alpha}i}
		-
		i \frac{1}{2}T \delta^\alpha_\beta \bar{\epsilon}^{\dot{\alpha}i}
		-
		i \frac{1}{2}\bar{P}^{\dot{\alpha}}_\beta \epsilon^{\alpha i}
		+
		\frac{1}{6} \epsilon_{\beta j} V^{\alpha\dot{\alpha}(ij)}
		\\&
		+
		\psi^{\rho\dot{\rho}}_{\beta} \partial_{\rho\dot{\rho}} a^{\alpha\dot{\alpha}}
		- a^{\rho\dot{\rho}} \partial_{\rho\dot{\rho}} \psi^{\alpha\dot{\alpha}i}_\beta
		-
		l_{(\beta}^{\;\;\rho)} \psi_\rho^{\alpha\dot{\alpha}i}
		,
	\end{split}
\end{equation}
\begin{equation}
	\delta C_{\alpha\dot{\alpha}} =
	(\delta_{\alpha}^\rho\delta_{\dot{\alpha}}^{\dot{\rho}} + \Phi_{\alpha\dot{\alpha}}^{\rho\dot{\rho}}) \partial_{\rho\dot{\rho}} c
	-
	a^{\rho\dot{\rho}} \partial_{\rho\dot{\rho}} C_{\alpha\dot{\alpha}}
	-
	(l^{\;\;\beta)}_{(\alpha}\delta^{\dot{\beta}}_{\dot{\alpha}} + \delta_\alpha^\beta l_{(\dot{\alpha}}^{\;\;\dot{\beta})}) C_{\beta\dot{\beta}}
	- i \epsilon_{\beta i} \bar{\rho}^i_{\dot{\beta}}
	- i \epsilon_{\dot{\beta} i} \rho^i_{\beta}.
\end{equation}
\end{subequations}

These transformations make it possible to link the harmonic superspace  formulation to the component approach. For example, we see that the combination
$\delta_\beta^{\alpha}\delta^{\dot{\alpha}}_{\dot{\beta}}+\Phi^{\alpha\dot{\alpha}}_{\beta\dot{\beta}} := \frac{1}{2}\sigma^{\alpha\dot{\alpha}}_m \sigma^a_{\beta\dot{\beta}} e_a^m$ can be identified with the fierbein
$e^m_a$, the parameter $a^{\alpha\dot{\alpha}}$ stands for local translations and $l^{(\alpha\beta)}, \bar{l}^{(\dot{\alpha}\dot{\beta})}$ for  local Lorentz transformations, while $\epsilon^{\alpha i},
\bar{\epsilon}^{\dot{\alpha} i}$ are the parameters of local supersymmetry. The metric tensor is then given by $g^{mn} = e^m_a e^n_b \eta^{ab}$ and have the standard transformation law $\delta g^{mn} = \partial_k a^m g^{kn}
+ \partial_k a^n g^{mk} - a^k \partial_k g^{mn}$. The fermionic field $\psi_\beta^{\alpha\dot{\alpha}i}$ describes the doublet of the spin $\frac{3}{2}$ Rarita-Schwinger fields,  $C_{\alpha\dot{\alpha}}$ is the spin 1 gauge
field. Other fields are auxiliary fields of the off-shell $\mathcal{N}=2$ Einstein supergravity multiplet\footnote{Some auxiliary fields require redefinition, see the explicit redefinition in the linearized limit \eqref{eq:
	field redefinition}.}:
	\begin{equation}
\begin{split}
	\text{Physical sector}&: \qquad\Phi^{\alpha\dot{\alpha}}_{\beta\dot{\beta}}, \psi^{\alpha\dot{\alpha}i}_{\beta}, C_{\beta\dot{\beta}};
	\\
	\text{Auxiliary sector}&:
	\qquad V^{\alpha\dot{\alpha}(ij)}, \rho^i_\beta, S^{(ij)}, T,  T^{(\alpha\beta)},P^{\alpha\dot{\beta}}, \chi^{\alpha i}.
\end{split}
\end{equation}

Totally, the multiplet carries $\mathbf{40_B + 40_F}$ off-shell degrees of freedom.
The dimensions of the involved fields can be directly determined from that of the prepotentials in WZ gauge  \eqref{eq: WZ gauge}.
The above component WZ expansion corresponds to the ``minimal'' version of $\mathcal{N} = 2$ supergravity obtained from the conformal one
through the compensation by the so called ``nonlinear multiplet'' (in our formulation there are in fact two hidden compensators: $\mathcal{N}=2$ vector multiplet and the nonlinear multiplet,
see, e.g., \cite{18,Freedman:2012zz, Lauria:2020rhc}). The off-shell component content of this version of $\mathcal{N}=2$ Einstein supergravity
was  firstly found by Fradkin and Vasiliev  \cite{Fradkin:1979cw, Fradkin:1979as} and, independently, by de Wit and van Holten \cite{deWit:1979xpv} .

\subsection{$\mathcal{N}=2$ supergravity multiplet: linearization}

Linearizing gauge transformations \eqref{eq: perp GF} yields:
\begin{equation}\label{eq: lin GF}
\begin{cases}
	\delta_\lambda h^{++\alpha\dot{\alpha}}
	=
	\mathcal{D}^{++} \lambda^{\alpha\dot{\alpha}}
	+4i \lambda^{+\alpha} \bar{\theta}^{+\dot{\alpha}}
	+ 4i \theta^{+\alpha} \bar{\lambda}^{+\dot{\alpha}},
	\\
	\delta_\lambda h^{++\hat{\alpha}+} = \mathcal{D}^{++} \lambda^{+\hat{\alpha}},
	\\
	\delta_\lambda h^{++\hat{\alpha}-} = \mathcal{D}^{++}\lambda^{-\hat{\alpha}}
	-
	\lambda^{+\hat{\alpha}},
	\\
	\delta_\lambda h^{++5} = \mathcal{D}^{++}\lambda^5 - 2i \lambda^{+\hat{\alpha}} \theta^+_{\hat{\alpha}}.
\end{cases}
\end{equation}
In the linearized case we also impose the analytic gauge $h^{++\hat{\alpha}-}=0$, so $\mathcal{D}^{++}\lambda^{-\hat{\alpha}} = \lambda^{+\hat{\alpha}}$.
In this gauge all prepotentials are analytic. The WZ gauge has the same form as in the nonlinear theory, eq.  \eqref{eq: WZ gauge}.

Rigid supersymmetry transformations of the analytic prepotentials  can be obtained by requiring invariance of the covariant derivative $\mathfrak{D}^{++}$ under the rigid $\mathcal{N}=2$ supersymmetry:
\begin{equation}\label{eq: SUSY D++}
\delta_\epsilon \mathfrak{D}^{++} = [\hat{\Lambda}_{\epsilon}, \mathcal{D}^{++}] + \delta_\epsilon h^{++M} \partial_M = 0.
\end{equation}
As a solution of \eqref{eq: SUSY D++}, all parameters $\lambda_\epsilon^M$ span rigid supersymmetry:
\begin{equation}
\lambda_\epsilon^{\alpha\dot{\alpha}} = -4i \theta^{+\alpha} \bar{\epsilon}^{-\dot{\alpha}}
-
4i \epsilon^{-\alpha} \bar{\theta}^{+\dot{\alpha}},
\quad
\lambda_\epsilon^{+\alpha} = \epsilon^{+\alpha},
\quad
\lambda_\epsilon^{+\dot{\alpha}} = \bar{\epsilon}^{+\dot{\alpha}},
\quad
\lambda_\epsilon^5 = 2i \epsilon^{-\hat{\alpha}} \theta^+_{\hat{\alpha}}.
\end{equation}
Note that, in full agreement with \eqref{eq: SUSY transf}, $\delta_\epsilon z^M = \lambda_\epsilon^M$.

From \eqref{eq: SUSY D++} we find that the rigid supersymmetry transformations of the prepotentials are given by:
\begin{equation}\label{eq: N=2 susy}
\begin{cases}
	\delta_\epsilon h^{++\alpha\dot{\alpha}} = -4i h^{++\alpha+} \bar{\epsilon}^{-\dot{\alpha}}
	- 4i \epsilon^{-\alpha}h^{++\dot{\alpha}+}
	-
	\delta_\epsilon z^M \partial_M h^{++\alpha\dot{\alpha}},
	\\
	\delta_\epsilon h^{++\alpha+} =
	- \delta_\epsilon z^M \partial_M\delta_\epsilon h^{++\alpha+},
	\\
	\delta_\epsilon h^{++\dot{\alpha}+} =
	- \delta_\epsilon z^M \partial_M  h^{++\dot{\alpha}+},
	\\
	\delta_\epsilon h^{++5} = 2i h^{++\hat{\alpha}+} \epsilon^-_{\hat{\alpha}} - \delta_\epsilon z^M \partial_M h^{++5}.
\end{cases}
\end{equation}
Note that the prepotentials $h^{++\alpha\dot{\alpha}}$ and $h^{++5}$ have non-standard transformation laws. As usual, in WZ gauge, supersymmetry
transformations must be accompanied by compensating gauge transformations that ensure the preservation of the gauge-fixed form of $h^{++M}_{WZ}$.

From \eqref{eq: N=2 susy} one can deduce supersymmetry transformations of the component fields. For example, the gravity field transforms as:
\begin{equation}
\delta_\epsilon \Phi^{\alpha\dot{\alpha}}_{\beta\dot{\beta}}
=
2i \left( \epsilon^i_\beta \bar{\psi}^{\alpha\dot{\alpha}}_{\dot{\beta}\;\;\;i} - \bar{\epsilon}^i_{\dot{\beta}} \psi^{\alpha\dot{\alpha}}_{\beta\;\;\;i}\right).
\end{equation}
The complete  set of linearized gauge transformation laws for the component fields can be deduced from  \eqref{eq: lin GF} and the linearized version of the  residual gauge parameters \eqref{WZ res}:
\begin{subequations}\label{eq: field gauge transf}
\begin{equation}\label{eq: 3.19a}
	\delta_\lambda \Phi^{\alpha\dot{\alpha}}_{\beta\dot{\beta}} = \partial_{\beta\dot{\beta}} a^{\alpha\dot{\alpha}} - l^{\;\;\alpha)}_{(\beta} \delta^{\dot{\alpha}}_{\dot{\beta}}
	-
	\delta^{\alpha}_\beta l^{\;\;\dot{\alpha})}_{(\dot{\beta}},
\end{equation}
\begin{equation}\label{eq: 3.19b}
	\delta_\lambda \psi_\beta^{\alpha\dot{\alpha}i}
	=
	-
	\partial^{\dot{\alpha}}_\beta \epsilon^{\alpha i},
\end{equation}
\begin{equation}\label{eq: 3.19c}
	\delta_\lambda C_{\alpha\dot{\alpha}} = \partial_{\alpha\dot{\alpha}} c,
\end{equation}
\begin{equation}\label{eq: P gauge transf}
	\delta_\lambda P^{\alpha\dot{\alpha}} = -2i \partial^{\dot{\alpha}\beta} l_{(\beta}^{\;\;\alpha)},
\end{equation}
\begin{equation}\label{eq: 3.19e}
	\delta_\lambda \chi^{\alpha i} = 2 \Box \epsilon^{\alpha i},
\end{equation}
\begin{equation}\label{eq: 3.19f}
	\delta_\lambda \rho_\alpha^i = \partial_{\alpha\dot{\alpha}} \bar{\epsilon}^{\dot{\alpha}i}.
\end{equation}
\end{subequations}
Other fields in WZ gauge are invariant under gauge transformations. The transformations \eqref{eq: 3.19a}, \eqref{eq: 3.19b} and \eqref{eq: 3.19c} are linearized gauge transformations for the spin $2$, spin $\frac{3}{2}$
and spin $1$ fields, respectively. The fields $P^{\alpha\dot{\alpha}},  \chi^{\alpha i}$ and $\rho_\alpha^i$ can be properly redefined with the help of the spin $2$ and spin $\frac{3}{2}$ gauge fields so as to ensure trivial
gauge transformation laws for them. Such redefinitions play an important role in the component reduction, so we will explicitly present them here.

\medskip

Spin 2 field has the following decomposition into the irreducible parts:
\begin{equation}
\Phi^{\alpha\dot{\alpha}\beta\dot{\beta}} = \Phi^{(\alpha\beta)(\dot{\alpha}\dot{\beta})} + \epsilon^{\alpha\beta} \Phi^{(\dot{\alpha}\dot{\beta})}
+
\epsilon^{\dot{\alpha}\dot{\beta}} \Phi^{(\alpha\beta)}
+
\epsilon^{\alpha\beta}\epsilon^{\dot{\alpha}\dot{\beta}} \Phi.
\end{equation}
Using local Lorentz transformations, one can fix the ``symmetric'' gauge: $\Phi^{(\alpha\beta)}=0$, $\Phi^{(\dot{\alpha}\dot{\beta})} =0$,
thus expressing local Lorentz transformations in terms of local translations $a^{\alpha\dot{\alpha}}$:
\begin{equation}\label{eq: local Lorentz fixed}
l^{\;\;\alpha)}_{(\beta} = \frac{1}{2} \partial_{(\beta\dot{\beta}} a^{\alpha)\dot{\beta}},
\qquad
l^{\;\;\dot{\alpha})}_{(\dot{\beta}} = \frac{1}{2} \partial_{\beta(\dot{\beta}}a^{\beta\dot{\alpha})}.
\end{equation}
In this gauge, the remaining components undergo the transformation laws:
\begin{equation}\label{eq: spin 2 lin}
\delta \Phi^{(\alpha\beta)(\dot{\alpha}\dot{\beta})} = \partial^{(\alpha(\dot{\alpha}} a^{\beta)\dot{\beta})},
\qquad
\delta \Phi = \frac{1}{4}\partial_{\beta\dot{\beta}}a^{\beta\dot{\beta}}.
\end{equation}

The field $P^{\alpha\dot{\alpha}}$ possesses the non-trivial gauge transformation law \eqref{eq: P gauge transf}. Because of the relations \eqref{eq: local Lorentz fixed}, in the symmetric gauge we have:
\begin{equation}
\partial^{\dot{\alpha}\beta} l_{(\beta}^{\;\;\alpha)}
=
\frac{1}{16} \Box a^{\alpha\dot{\alpha}}
-
\frac{1}{4} \partial^\alpha_{\dot{\beta}} \partial^{\dot{\alpha}}_{\beta} a^{\beta\dot{\beta}}
=
\frac{1}{8} \Box a^{\alpha\dot{\alpha}}
-
\frac{1}{4} \partial^{\alpha\dot{\alpha}} \partial^{\beta\dot{\beta}} a_{\beta\dot{\beta}}.
\end{equation}
Then the variation \eqref{eq: P gauge transf} can be canceled by adding to $P^{\alpha\dot{\alpha}}$ the term $iB^{\alpha\dot{\alpha}}$, where:
\begin{equation}
B^{\alpha\dot{\alpha}} =
- \partial_{\beta\dot{\beta}} \Phi^{(\alpha\beta)(\dot{\alpha}\dot{\beta})}
+3
\partial^{\alpha\dot{\alpha}} \Phi , \quad \delta B^{\alpha\dot\alpha} = -2 \partial^{\dot{\alpha}\beta} l_{(\beta}^{\;\;\alpha)}
=
- \frac{1}{4} \Box a^{\alpha\dot{\alpha}}
+
\frac{1}{2} \partial^{\alpha\dot{\alpha}} \partial^{\beta\dot{\beta}} a_{\beta\dot{\beta}}.
\end{equation}
Note the important property, $\partial_{\beta\dot{\beta}} B^{\beta\dot{\beta}} = - \frac{1}{4}R$, where $R$ is the linearized scalar curvature defined in \eqref{eq: R}.
It is invariant under the linearized gauge transformations \eqref{eq: spin 2 lin}.

\medskip

The spin $\frac{3}{2}$ field  can also be decomposed into the irreducible pieces:
\begin{equation}\label{eq: spin 3/2 irred parts}
\psi^{\beta\alpha\dot{\alpha}i} =   \psi^{(\beta\alpha)\dot{\alpha}i} + \epsilon^{\beta\alpha} \psi^{\dot{\alpha}i},
\qquad
\delta_\lambda \psi^{(\beta\alpha)\dot{\alpha}i} = - \partial^{(\beta\dot{\alpha}}\epsilon^{\alpha)i},
\quad
\delta_\lambda \psi^{\dot{\alpha}i} = -\frac{1}{2} \partial^{\dot{\alpha}}_\alpha \epsilon^{\alpha i}.
\end{equation}
The fields $\chi^{\alpha i}$ and $\rho_\alpha^i$ have non-trivial gauge transformation laws \eqref{eq: 3.19e} and \eqref{eq: 3.19f}.
So they need to be redefined in a similar way.

Finally, the complete set of the field redefinitions required amounts to:
\begin{equation}\label{eq: field redefinition}
P^{\alpha\dot{\alpha}} \to P^{\alpha\dot{\alpha}} +i B^{\alpha\dot{\alpha}},
\qquad
\chi^{\alpha i} \to \chi^{\alpha i} + 4 \partial^\beta_{\dot{\beta}} \psi_\beta^{\alpha\dot{\beta}i},
\qquad
\rho^i_\alpha \to \rho^i_\alpha - 2 \bar{\psi}^i_\alpha.
\end{equation}
After these redefinitions, the fields $P^{\alpha\dot{\alpha}}$, $\chi^{\alpha i}$ and $\rho_\alpha^i$ become invariant under gauge transformations and so can be treated as genuine auxiliary fields.

\section{$\mathcal{N}=2$ supergravity actions: nonlinear and linearized}\label{sec:4}

This section is a review of the construction of the nonlinear $\mathcal{N}=2$ supergravity action and its linearized version.
An important ingredient of the construction is the non-analytic negatively charged gauge potentials related to the fundamental analytic ones by the harmonic zero-curvature conditions.

\subsection{Nonlinear theory}

The action of $\mathcal{N}=2$ Einstein supergravity in harmonic superspace was constructed by Galperin, Ky and Sokatchev \cite{Galperin:1987ek}.
Here we briefly recall its main features. The action describes $\mathcal{N}=2$ abelian vector multiplet in the $\mathcal{N}=2$ supergravity
background\footnote{Integration over Grassmann variables in defined as: $d^8\theta = d^4\theta^+ d^4 \theta^- := (\mathcal{D}^-)^4(\mathcal{D}^+)^4$,
$(\mathcal{D}^{\pm})^4: = \frac{1}{16} (\mathcal{D}^+)^2 (\bar{\mathcal{D}}^+)^2$. Integral over harmonic variables is defined as $\int du \, 1 = 1$ and zero otherwise.}:
\begin{equation}\label{eq: N=2 sugra full}
S_{full} = - \int d^4x d^8\theta du \, E \, H^{++5} H^{--5}.
\end{equation}
Here $E$ is the $\mathcal{N}=2$ full harmonic superspace integration measure with the transformation law:
\begin{equation}
\delta_\lambda E = - \left( \partial_{\alpha\dot{\alpha}} \lambda^{\alpha\dot{\alpha}} - \partial^-_{\hat{\alpha}} \lambda^{+\hat{\alpha}} - \partial^+_{\hat{\alpha}} \lambda^{-\hat{\alpha}} \right) E.
\end{equation}

The measure $E$ can be constructed from supergravity prepotentials and it satisfies some useful properties. We shall not present the whole construction here, for details see \cite{Galperin:1987ek, 18, Ivanov:2022vwc}. It
will be important that it is expressed in terms of the potentials $H^{++M}$ and $H^{--M}$ which covariantize the harmonic derivatives ${\cal D}^{\pm\pm}$
\begin{subequations}
\begin{equation}
	\mathfrak{D}^{++} = \partial^{++} + H^{++\alpha\dot{\alpha}}\partial_{\alpha\dot{\alpha}} + H^{++\hat{\alpha}+}\partial^-_{\hat{\alpha}} + \theta^{+\hat{\alpha}}\partial^+_{\hat{\alpha}}
	+
	H^{++5} \partial_5,
\end{equation}
\begin{equation}
	\mathfrak{D}^{--} = \partial^{--} + H^{--\alpha\dot{\alpha}}\partial_{\alpha\dot{\alpha}} + H^{--\hat{\alpha}+}\partial^-_{\hat{\alpha}} + H^{--\hat{\alpha}-}\partial^+_{\hat{\alpha}}
	+
	H^{--5} \partial_5.
\end{equation}
\end{subequations}
The non-analytic potentials $H^{--M}$ are uniquely determined in terms of $H^{++M}$ as solutions of the harmonic zero-curvature condition:
\begin{equation}
[\mathfrak{D}^{++}, \mathfrak{D}^{--}] = D^0.
\end{equation}

One could also consider another nonlinear action:
\begin{equation} \label{ZeroAct}
S_{alt} = \int d^4x d^8\theta du \, E\, H^{++\alpha\dot{\alpha}} H^{--}_{\alpha\dot{\alpha}}.
\end{equation}
However, it vanishes due to the property (easily checkable in WZ gauge):
\begin{equation}
(\mathcal{D}^+)^4 \left( E H^{--\alpha\dot{\alpha}} \right) = 0.
\end{equation}

In this paper we limit our attention to studying invariants and their component reductions in the linear approximations, so we shall not discuss
any higher-curvature generalization of \eqref{eq: N=2 sugra full} or \eqref{ZeroAct}.

\subsection{Linearized action and covariant superfields}

To construct the linearized $\mathcal{N}=2$ supergravity action with manifest $\mathcal{N}=2$ supersymmetry  it is natural to pass to  the
covariant superfields, since the analytic prepotentials defined above have non-standard transformations laws \eqref{eq: N=2 susy} under rigid ${\cal N}=2$ supersymmetry.
The most natural way to introduce $\mathcal{N}=2$ covariant superfields is to  use the basis of covariant derivatives. To this end, one needs to split the covariant harmonic derivative
into the flat and the supergravity parts:
\begin{equation}
\mathfrak{D}^{++} = \mathcal{D}^{++} + \hat{\mathcal{H}}^{++},
\qquad
\hat{\mathcal{H}}^{++}: = h^{++M}\partial_M.
\end{equation}
These parts are independently invariant under rigid supersymmetry: $\delta_{\epsilon} \mathcal{D}^{++}=0$, $\delta_{\epsilon} \hat{\mathcal{H}}^{++}=0$.
As the next step, one rewrites the  supergravity  part $\hat{\mathcal{H}}^{++}$ in  the basis of covariant derivatives \eqref{eq: spinor cov}:
\begin{equation}
\hat{\mathcal{H}}^{++} = G^{++\alpha\dot{\alpha}} \partial_{\alpha\dot{\alpha}}
+
G^{++\alpha+} \mathcal{D}^-_\alpha
+
G^{++\dot{\alpha}+} \bar{\mathcal{D}}^-_{\dot{\alpha}}
+
G^{++5} \partial_5.
\end{equation}
Due to the invariance of both $\hat{\mathcal{H}}^{++}$ and covariant derivatives under rigid ${\cal N}=2$ supersymmetry, the new vielbeins $G^{++M}$
are $\mathcal{N}=2$ superfields with the standard rigid supersymmetry laws. However, most of the newly introduced superfields $G^{++M}$ are non-analytic:
\begin{equation}
\begin{split}
	&G^{++\alpha\dot{\alpha}} = h^{++\alpha\dot{\alpha}} + 4i h^{++\alpha+} \bar{\theta}^{-\dot{\alpha}}
	+
	4i \theta^{-\alpha} h^{++\dot{\alpha}+},
	\\
	&G^{++5} \;\,\,= h^{++5} -2i h^{++\hat{\alpha}+} \theta^{-}_{\hat{\alpha}},
	\\
	&G^{++\hat{\alpha}+} = - h^{++\hat{\alpha}+}.
	\\
	&G^{++\hat{\alpha}-} = h^{++\hat{\alpha}-} = 0.
\end{split}
\end{equation}

The linearized gauge transformations of $\hat{\mathcal{H}}^{++}$ are given by:
\begin{equation}
\delta_\lambda \hat{\mathcal{H}}^{++} = [\mathcal{D}^{++}, \hat{\Lambda}],
\qquad
\hat{\Lambda}:= \lambda^M \partial_M = \Lambda^M \mathcal{D}_M,
\end{equation}
where the parameters $\Lambda^M$ are defined as:
\begin{equation}
\begin{split}
	&\Lambda^{\alpha\dot{\alpha}} = \lambda^{\alpha\dot{\alpha}} + 4i \lambda^{+\alpha} \bar{\theta}^{-\dot{\alpha}} + 4i \theta^{-\alpha} \lambda^{+\dot{\alpha}},
	\\
	&\Lambda^5 \;\;\,= \lambda^5 - 2i \lambda^{+\hat{\alpha}} \theta^-_{\hat{\alpha}},
	\\
	&\Lambda^{+\hat{\alpha}} = - \lambda^{+\hat{\alpha}},
	\\
	&\Lambda^{-\hat{\alpha}} = \lambda^{-\hat{\alpha}},
	\qquad \mathcal{D}^{++}\Lambda^{-\hat{\alpha}} + \Lambda^{+\hat{\alpha}} = 0.
\end{split}
\end{equation}
The linearized gauge transformations of the covariant superfields $G^{++M}$ are then obtained in the form:
\begin{equation}
\delta_\lambda G^{++\alpha\dot{\alpha}} = \mathcal{D}^{++} \Lambda^{\alpha\dot{\alpha}},
\quad
\delta_\lambda G^{++5} = \mathcal{D}^{++} \Lambda^5,
\quad
\delta_\lambda G^{++\hat{\alpha}+} = \mathcal{D}^{++} \Lambda^{+\hat{\alpha}}.
\end{equation}
The linearized zero-curvature condition reads:
\begin{equation}\label{eq: zero curv lin}
[\mathcal{D}^{++}, \hat{\mathcal{H}}^{--}] = [\mathcal{D}^{--}, \hat{\mathcal{H}}^{++}].
\end{equation}
Here we have introduced the negatively charged differential operator $\hat{\mathcal{H}}^{--}$:
\begin{equation}
\begin{aligned}
	&\mathfrak{D}^{--} := \mathcal{D}^{--} + \hat{\mathcal{H}}^{--},
	\\
	&\hat{\mathcal{H}}^{--} := G^{--\alpha\dot{\alpha}}\partial_{\alpha\dot{\alpha}}
	+
	G^{--\hat{\alpha}+} \mathcal{D}^-_{\hat{\alpha}}
	+
	G^{--\hat{\alpha}-} \mathcal{D}^+_{\hat{\alpha}}
	+
	G^{--5} \partial_5.
\end{aligned}
\end{equation}
By construction, it trivially transforms under rigid $\mathcal{N}=2$ supersymmetry, $\delta_\epsilon \hat{\mathcal{H}}^{--}=0$.

In terms of the covariant superfields the zero-curvature condition \eqref{eq: zero curv lin} amounts to the set of equations:
\begin{equation}\label{eq: zero curv eq comp}
\begin{cases}
	\mathcal{D}^{++} G^{--\alpha\dot{\alpha}}
	=
	\mathcal{D}^{--} G^{++\alpha\dot{\alpha}},
	\\
	\mathcal{D}^{++} G^{--5} = \mathcal{D}^{--} G^{++5},
	\\
	\mathcal{D}^{++}G^{--\hat{\alpha}+}
	=
	\mathcal{D}^{--} G^{++\hat{\alpha}+},
	\\
	\mathcal{D}^{++} G^{--\hat{\alpha}-}
	+
	G^{--\hat{\alpha}+}
	=0.
\end{cases}
\end{equation}
These are linear harmonic equations and they have a unique solution \cite{18}. Some constituents of this solution are explicitly quoted in Appendix \ref{app: zero curv solutions}.

The zero-curvature condition \eqref{eq: zero curv lin} is invariant under the linearized gauge transformations:
\begin{equation}
\delta_\lambda \hat{\mathcal{H}}^{++} = [\mathcal{D}^{++}, \hat{\Lambda}],
\qquad
\delta_\lambda \hat{\mathcal{H}}^{--} = [\mathcal{D}^{--}, \hat{\Lambda}].
\end{equation}
Then the linearized gauge freedom of the potentials $G^{--M}$ is given by:
\begin{equation}\label{eq: zero-curv G}
\begin{split}
	\delta_\lambda G^{--\alpha\dot{\alpha}} = \mathcal{D}^{--} \Lambda^{\alpha\dot{\alpha}},
	\quad
	\delta_\lambda G^{--5} = \mathcal{D}^{--} \Lambda^5,
	\quad
	\\
	\delta_\lambda G^{--\hat{\alpha}+} = \mathcal{D}^{--} \Lambda^{+ \hat{\alpha}} + \Lambda^{-\hat{\alpha}},
	\quad
	\delta_\lambda G^{--\hat{\alpha}-} = \mathcal{D}^{--} \Lambda^{-\hat{\alpha}}.
\end{split}
\end{equation}

Now it is not difficult to verify that the supersymmetric and gauge invariant quadratic action has the form:
\begin{equation}\label{eq: N=2 lin Einstein}
S_{lin} = - \int d^4x d^8 \theta du \left[ G^{++\alpha\dot{\alpha}} G^{--}_{\alpha\dot{\alpha}} + 4G^{++5} G^{--5} \right].
\end{equation}
Indeed, the gauge transformation of this action is:
\begin{equation} \label{VarFullS}
\delta_\lambda S_{lin} =  - 2\int d^4x d^8 \theta du
\left[ G^{++\alpha\dot{\alpha}} \mathcal{D}^{--} \Lambda_{\alpha\dot{\alpha}} + 4G^{++5} \mathcal{D}^{--} \Lambda^{5} \right].
\end{equation}
Integrating over non-analytic Grassmann variables $d^4\theta^- := (\mathcal{D}^+)^4 = \frac{1}{16} (\mathcal{D}^+)^2 (\bar{\mathcal{D}}^+)^2$ yields:
\begin{equation}
(\mathcal{D}^+)^4 \left( G^{++\alpha\dot{\alpha}} \mathcal{D}^{--} \Lambda_{\alpha\dot{\alpha}} \right)
=
16i h^{++\alpha+} \partial_{\alpha\dot{\alpha}} \lambda^{+\dot{\alpha}}
-16i h^{++\dot{\alpha}+} \partial_{\alpha\dot{\alpha}} \lambda^{+\alpha},
\end{equation}
\begin{equation}
(\mathcal{D}^+)^4 \left( G^{++5} \mathcal{D}^{--} \Lambda^4 \right)
=
-4i h^{++\alpha+} \partial_{\alpha\dot{\alpha}} \lambda^{+\dot{\alpha}}
+4i h^{++\dot{\alpha}+} \partial_{\alpha\dot{\alpha}} \lambda^{+\alpha}.
\end{equation}
We observe that these terms in \eqref{VarFullS} explicitly cancel each other.

In section \ref{sec:6} we shall perform a detailed analysis of the component structure of this action
and explicitly demonstrate that it indeed describes the linearized off-shell $\mathcal{N}=2$ supergravity.

\section{Linearized super-curvatures and quadratic invariants}\label{sec:5}\label{sec: QI}

In this section, using the covariant superfields $G^{\pm\pm M}$ introduced in the previous section, we construct the linearized supercurvatures which contain linearized gravity invariants $R$,
$\mathcal{R}_{(\alpha\beta)(\dot{\alpha}\dot{\beta})}$, $\mathcal{R}_{(\alpha\beta\gamma\delta)}$, $\bar{\mathcal{R}}_{(\dot{\alpha}\dot{\beta}\dot{\gamma}\dot{\delta})}$ (see Appendix \ref{sec:A1} for notations).
We shall present all the superfield actions which in components involve terms quadratic in the gravity curvatures.

Inspecting the gauge transformations of the negatively charged  potentials \eqref{eq: zero-curv G} and generalizing the method which was exploited for construction of the linearized $\mathcal{N}=2$
supergravity action, we can construct two manifestly supersymmetric analytic objects:
\begin{subequations}
\begin{equation}\label{eq: SC1}
	\mathcal{F}^{++\alpha\dot{\alpha}} := \left( \mathcal{D}^+\right)^4 G^{--\alpha\dot{\alpha}},
\end{equation}
\begin{equation}\label{eq: SC2}
	\mathcal{F}^{++5} := \left( \mathcal{D}^+\right)^4 G^{--5}.
\end{equation}
\end{subequations}
These quantities are invariant under the linearized gauge transformations \eqref{eq: zero-curv G}:
\begin{subequations}
\begin{equation}
	\delta_\lambda \mathcal{F}^{++\alpha\dot{\alpha}}
	=
	(\mathcal{D}^+)^4 \delta G^{--\alpha\dot{\alpha}}
	=
	(\mathcal{D}^+)^4 \mathcal{D}^{--} \Lambda^{\alpha\dot{\alpha}} = 0,
\end{equation}
\begin{equation}
	\delta_\lambda \mathcal{F}^{++5}
	=
	(\mathcal{D}^+)^4 \delta G^{--5}
	=
	(\mathcal{D}^+)^4 \mathcal{D}^{--} \Lambda^{5} = 0.
\end{equation}
\end{subequations}
It is then natural to treat these quantities as linearized $\mathcal{N}=2$ supergravity curvatures.
Note that, as a consequence of the superfield equations of motion of $\mathcal{N}=2$ linearized Einstein supergravity \eqref{eq: N=2 lin Einstein}  \cite{Buchbinder:2022vra},
these supercurvatures are zero on shell of the latter theory:
\begin{equation}
\mathcal{F}^{++\alpha\dot{\alpha}} = 0, \qquad \mathcal{F}^{++5} = 0.
\end{equation}

Using these linearized supercurvatures, one can build two $\mathcal{N}=2$ linearized supergravity invariants
\begin{equation}\label{eq: inv 1}
I_1
:=
\int d\zeta^{(-4)} \mathcal{F}^{++\alpha\dot{\alpha}} \mathcal{F}^{++}_{\alpha \dot{\alpha}}
= \int d^4x d^8 \theta du \;    G^{--\alpha\dot{\alpha}} (\mathcal{D}^+)^4 G^{--}_{\alpha\dot{\alpha}} ,
\end{equation}

\begin{equation}\label{eq: inv 2}
I_2
:=
\int d\zeta^{(-4)} \mathcal{F}^{++5} \mathcal{F}^{++5}
= \int d^4x d^8\theta du \; G^{--5} (\mathcal{D}^+)^4 G^{--5}
.
\end{equation}
In the second expressions the integration goes  over the total harmonic superspace.
The invariants \eqref{eq: inv 1} and \eqref{eq: inv 2} have the same form as the invariant constructed in \cite{Ivanov:2005qf} in six-dimensional renormalizable higher-derivative
gauge theory and so can be considered as its spin 2 generalization. As we shall see in the next section, at the component level
these invariants provide $\mathcal{N}=2$ supergravity extensions of
the linearized invariants $\frac{1}{4}R^{mn} R_{mn} = 16 \mathcal{R}^{(\alpha\beta)(\dot{\alpha}\dot{\beta})}\mathcal{R}_{(\alpha\beta)(\dot{\alpha}\dot{\beta})} + R^2$ and $R^2$,  respectively.
\medskip

It is also possible to construct one more quadratic invariant, with a more complicated structure.
First, we define the object:
\begin{equation}\label{eq: mathcal H}
\mathcal{H}^{++}_{\alpha\beta} := \mathcal{D}^-_\alpha G^{+++}_\beta + \partial_\alpha^{\dot{\rho}} G^{++}_{\beta\dot{\rho}}.
\end{equation}
Since it is constructed out of the covariant superfields, it is manifestly supersymmetric.
Under the linearized gauge transformation it behaves as:
\begin{equation}
\delta_\lambda \mathcal{H}^{++}_{\alpha\beta} = \mathcal{D}^{++} \Lambda_{\alpha\beta},
\qquad
\Lambda_{\alpha\beta} := \mathcal{D}^-_\alpha \Lambda^+_\beta 
+
 \partial_\alpha^{\dot{\rho}} \Lambda_{\beta\dot{\rho}}.
\end{equation}

The newly introduced object is \textit{half-analytic} :
\begin{equation}\label{eq: half-analiticity}
\bar{\mathcal{D}}^+_{\dot{\alpha}} \mathcal{H}^{++}_{\alpha\beta} = 0,
\qquad
\bar{\mathcal{D}}^+_{\dot{\alpha}} \Lambda_{\alpha\beta} = 0.
\end{equation}
This property fixes the relative coefficient in \eqref{eq: mathcal H}. The superfield $\mathcal{H}^{++}_{\alpha\beta}$ can be decomposed into the irreducible parts as
$\mathcal{H}^{++}_{\alpha\beta} = \mathcal{H}^{++}_{(\alpha\beta)} + \epsilon_{\alpha\beta} \mathcal{H}^{++}$.
Using the zero-curvature equation, one can define its negatively charged counterpart $\mathcal{H}^{--}_{\alpha\beta}$:
\begin{equation}\label{eq: weyl zc}
\mathcal{D}^{++} \mathcal{H}^{--}_{\alpha\beta}
=
\mathcal{D}^{--} \mathcal{H}^{++}_{\alpha\beta}.
\end{equation}
The superfield $\mathcal{H}^{--}_{\alpha\beta} $ has the following  linearized gauge transformation law:
\begin{equation}
\delta_\lambda \mathcal{H}^{--}_{\alpha\beta} = \mathcal{D}^{--} \Lambda_{\alpha\beta}.
\end{equation}
Using the zero-curvature equations \eqref{eq: zero curv eq comp} one can explicitly express $\mathcal{H}^{--}_{\alpha\beta}$ in terms of the negatively charged potentials $G^{--M}$:
\begin{equation}
\mathcal{H}^{--}_{\alpha\beta} = \mathcal{D}^+_\alpha G^{---}_\beta
+
\mathcal{D}^-_\alpha G^{--+}_\beta
+
\partial_\alpha^{\dot{\rho}}G^{--}_{\beta\dot{\rho}}.
\end{equation}

With the help of the irreducible parts of $\mathcal{H}^{++}_{\alpha\beta} $ and $\mathcal{H}^{--}_{\alpha\beta} $ one can construct two invariant actions:
\begin{equation} \label{I3invA}
I_3 = \int d^4xd^8\theta du \; \mathcal{H}^{++(\alpha\beta)} \mathcal{H}^{--}_{(\alpha\beta)},
\end{equation}
\begin{equation}\label{eq I4}
I_4 = \int d^4xd^8\theta du \; \mathcal{H}^{++} \mathcal{H}^{--}.
\end{equation}
The invariance of $I_3$ and $I_4$ directly follows from the half-analyticity condition $\eqref{eq: half-analiticity}$. For example, the transformation of $I_3$ is given by:
\begin{equation}
\begin{split}
	\delta_\lambda I_3 &= 2 \int d^4x d^8\theta du\; \mathcal{H}^{++(\alpha\beta)} \mathcal{D}^{--} \Lambda_{(\alpha\beta)}
	\\&=
	\frac{1}{8}\int d^4x d^4 \theta^+ (\mathcal{D}^+)^2
	\mathcal{H}^{++(\alpha\beta)} (\bar{\mathcal{D}}^+)^2\mathcal{D}^{--} \Lambda_{(\alpha\beta)}.
\end{split}
\end{equation}
Then we can use the half-analyticity property \eqref{eq: half-analiticity},
\begin{equation}
(\bar{\mathcal{D}}^+)^2\mathcal{D}^{--} \Lambda_{(\alpha\beta)}
=
- \bar{\mathcal{D}}^+_{\dot{\alpha}} \bar{\mathcal{D}}^{-\dot{\alpha}} \Lambda_{(\alpha\beta)}
=
\bar{\mathcal{D}}^{-\dot{\alpha}} \bar{\mathcal{D}}^+_{\dot{\alpha}}  \Lambda_{(\alpha\beta)}
= 0.
\end{equation}

The invariants $I_3$ and $I_4$ have the same structure as the action of $\mathcal{N}=2$ vector multiplet \cite{18}. The gauge invariance
of the latter action was ensured by the analyticity of both the prepotential $V^{++}$ and the gauge parameter $\lambda$. In checking the invariance of  $I_3$ and $I_4$
we used the weaker condition of half-analyticity.\\

There is yet another interesting way to represent these invariants.
Let us define the gauge invariant superfield:
\begin{equation}\label{eq: SC3}
\mathcal{W}_{\alpha\beta} :=(\bar{\mathcal{D}}^+)^2 \mathcal{H}^{--}_{\alpha\beta},
\qquad\qquad
\delta_\lambda \mathcal{W}_{\alpha\beta}=0.
\end{equation}
As we shall show in the next section,   the tensor  $\mathcal{W}_{(\alpha\beta)}$ in components involves the linearized Weyl
tensor $\mathcal{R}_{(\alpha\beta\gamma\delta)}$, so it is natural to dub $\mathcal{W}_{(\alpha\beta)}$ as the linearized  $\mathcal{N}=2$ super-Weyl tensor.
Antisymmetric part $\mathcal{W}$ does not involve gravitational curvatures and so \eqref{eq I4} cannot lead to any gravitation quadratic invariant, as will be shown below.

The tensor \eqref{eq: SC3} satisfies some useful properties:
\begin{equation}\label{eq: Weyl prop}
\mathcal{D}^{++} \mathcal{W}_{\alpha\beta} = 0,
\qquad
\mathcal{D}^{--} \mathcal{W}_{\alpha\beta} = 0,
\qquad
\bar{\mathcal{D}}^+_{\dot{\alpha}} \mathcal{W}_{\alpha\beta} = 0,
\qquad
\bar{\mathcal{D}}^-_{\dot{\alpha}} \mathcal{W}_{\alpha\beta} = 0.
\end{equation}
The last property can be checked in the following way:
\begin{equation}
\mathcal{D}^{++} \bar{\mathcal{D}}^-_{\dot{\alpha}} \mathcal{W}_{\alpha\beta} = \bar{\mathcal{D}}^+_{\dot{\alpha}} \mathcal{W}_{\alpha\beta} +       \bar{\mathcal{D}}^-_{\dot{\alpha}}
\mathcal{D}^{++} \mathcal{W}_{\alpha\beta}= 0\quad
\Rightarrow
\quad \bar{\mathcal{D}}^-_{\dot{\alpha}} \mathcal{W}_{\alpha\beta} = 0.
\end{equation}
Here we used the proposition that a negatively charged harmonic object annihilated by the harmonic derivative $\mathcal{D}^{++}$ is zero.
So we conclude, that $\mathcal{W}_{\alpha\beta}$ is covariantly harmonic-independent chiral tensor.

Integrals over  Grassmann variables can be represented as  $d^8\theta := d^4\theta d^4 \bar{\theta} = \frac{1}{16} d^4 \theta (\bar{\mathcal{D}}^+)^2 (\bar{\mathcal{D}}^-)^2$,
so due to the half-analyticity condition for $\mathcal{H}^{++(\alpha\beta)}$, the invariant $I_3$ can be rewritten as:
\begin{equation} \label{I3invB}
I_3 = \frac{1}{16} \int d^4x d^4\theta du\; (\bar{\mathcal{D}}^-)^2 \mathcal{H}^{++(\alpha\beta)} \mathcal{W}_{(\alpha\beta)}.
\end{equation}
Using the relation $\bar{\mathcal{D}}^{-\dot{\alpha}} = [\mathcal{D}^{--}, \bar{\mathcal{D}}^{+\dot{\alpha}}]$ and the properties of the linearized $\mathcal{N}=2$ super-Weyl tensor \eqref{eq: Weyl prop},
we obtain:
\begin{equation}
I_3 = \frac{1}{16} \int d^4x d^4\theta du \; \bar{\mathcal{D}}^-_{\dot{\alpha}} \mathcal{D}^{+\dot{\alpha}} \mathcal{D}^{--} \mathcal{H}^{++(\alpha\beta)} \mathcal{W}_{(\alpha\beta)}.
\end{equation}
Finally, using the zero-curvature condition     \eqref{eq: weyl zc} and integrating by parts, we find:
\begin{equation}\label{I3invC}
I_3 = - \frac{1}{16} \int d^4 x d^4\theta \; \mathcal{W}^{(\alpha\beta)} \mathcal{W}_{(\alpha\beta)}.
\end{equation}
Here we have also used the covariant harmonic-independence of $\mathcal{W}_{(\alpha\beta)}$.
So we end up with the integral over $\mathcal{N}=2$ chiral superspace.

Using the same reasonings, the invariant $I_4$ can also  be represented in the chiral form:
\begin{equation}
I_4 = - \frac{1}{16} \int d^4x d^4\theta \; \mathcal{W}^2.
\end{equation}

In such a form, the invariants $I_3$ and $I_4$ are manifestly gauge invariant.

Quite analogously, one can introduce the conjugated superfields $\bar{\mathcal{H}}^{++}_{(\dot{\alpha}\dot{\beta})}$,
$\bar{\mathcal{H}}^{--}_{(\dot{\alpha}\dot{\beta})}$, $\bar{\mathcal{W}}_{(\dot{\alpha}\dot{\beta})}$ and construct the conjugated invariants.

\section{Component contents of the linearized invariants}\label{sec:6}

Given the component expansion of the prepotentials in WZ gauge \eqref{eq: WZ gauge}, it is straightforward to deduce the component expansion of the invariants constructed.
Technically, the most involved part of the component reduction procedure is solving the harmonic  zero-curvature equations \eqref{eq: zero curv eq comp}.
The solutions of these equations are collected in Appendix \ref{app: zero curv solutions}. Using these results, in this section we present
component expansions of all invariants constructed  earlier. In some cases, if necessary, we present the full component structure, while otherwise quote only  the relevant terms  in the gravity sector.

\subsection{Einstein supergravity}

We start with the component reduction of the linearized $\mathcal{N}=2$ supergravity action \eqref{eq: N=2 lin Einstein}. It is convenient to consider the bosonic and fermionic parts of the action separately.

After substituting the solutions of the zero-curvature equations and integrating over Grassmann and harmonic co-ordinates, we obtain the linearized off-shell  $\mathcal{N}=2$ Einstein supergravity action in the bosonic sector
as:
\begin{equation} \label{eq: sugra bos0}
\begin{split}
	S^{(bos)}_{lin} = 8 \int d^4x \; \Bigr[ &
	\Phi^{\alpha\dot{\alpha}\beta\dot{\beta}} \mathcal{G}_{\alpha\dot{\alpha}\beta\dot{\beta}}
	+
	4 C^{\rho\dot{\rho}}\left(\partial_{\rho\dot{\rho}} \partial^{\beta\dot{\beta}}C_{\beta\dot{\beta}} - \frac{1}{2} \Box C_{\rho\dot{\rho}}  \right)
	\\&
	+
	4 i T^{(\alpha\beta)} \partial_{(\alpha}^{\dot{\beta}} C_{\dot{\beta}\beta)}
	-
	4 i \bar{T}^{(\dot{\alpha}\dot{\beta})} \partial_{(\dot{\alpha}}^{\beta} C_{\beta\dot{\beta})}
	\\&
	+ \frac{1}{2} T^{(\alpha\beta)} T_{(\alpha\beta)}
	+
	\frac{1}{2} T^{(\dot{\alpha}\dot{\beta})} T_{(\dot{\alpha}\dot{\beta})}
	-
	2 T\bar{T}
	\\&- \frac{1}{2} P^{\alpha\dot{\alpha}} \bar{P}_{\alpha\dot{\alpha}}
	- \frac{3}{4} P^{\alpha\dot{\alpha}} P_{\alpha\dot{\alpha}}
	-
	\frac{3}{4}
	\bar{P}^{\alpha\dot{\alpha}}
	\bar{P}_{\alpha\dot{\alpha}}
	\\&
	-
	\frac{1}{72}   V^{\alpha\dot{\alpha}(ij)}V_{\alpha\dot{\alpha}(ij)}
	-
	\frac{1}{18} S^{ij}S_{ij} \Bigr].
\end{split}
\end{equation}
It will be useful to decompose the electromagnetic tensor into the irreducible parts:
\begin{equation}
\mathcal{F}_{\rho\dot{\rho}\beta\dot{\beta}}
=
\partial_{\rho\dot{\rho}} C_{\beta\dot{\beta}}  - \partial_{\beta\dot{\beta}} C_{\rho\dot{\rho}}
=
\epsilon_{\rho\beta} \partial_{(\dot{\rho}}^\sigma C_{\sigma\dot{\beta})}
+
\epsilon_{\dot{\rho}\dot{\beta}} \partial_{(\rho}^{\dot{\sigma}} C_{\dot{\sigma} \beta)}
:=
\epsilon_{\rho\beta} \bar{\mathcal{F}}_{(\dot{\rho}\dot{\beta})}
+
\epsilon_{\dot{\rho}\dot{\beta}} \mathcal{F}_{(\rho \beta)}.
\end{equation}
The irreducible spin-tensors $\mathcal{F}_{(\rho \beta)}$ and $\bar{\mathcal{F}}_{(\dot{\rho}\dot{\beta})}$ are gauge invariant and they will be used for finding out
the  component structure of invariants.

For the Maxwell part of the action \eqref{eq: sugra bos0} we find (up to integration by parts):
\begin{equation}\label{eq: sugra bos}
\begin{split}
	32 C^{\rho\dot{\rho}}\left(\partial_{\rho\dot{\rho}} \partial^{\beta\dot{\beta}}C_{\beta\dot{\beta}} - \frac{1}{2} \Box C_{\rho\dot{\rho}}  \right)
	&=
	16 \left( \partial^{\beta\dot{\beta}}C^{\rho\dot{\rho}} - \partial^{\rho\dot{\rho}}C^{\beta\dot{\beta}}   \right)
	\left( \partial_{\beta\dot{\beta}}C_{\rho\dot{\rho}} - \partial_{\rho\dot{\rho}}C_{\beta\dot{\beta}}   \right)
	\\& = -32 \bar{\mathcal{F}}^{(\dot{\beta}\dot{\rho})}
	\bar{\mathcal{F}}_{(\dot{\beta}\dot{\rho})}
	-32 \mathcal{F}^{(\beta\rho)}
	\mathcal{F}_{(\beta\rho)}.
\end{split}
\end{equation}
This  Maxwell action has the incorrect sign. This is because, within the compensating procedure of deriving the Einstein supergravity action from the conformal supergravity one,
the above Maxwell field  comes from the vector superconformal compensator action with a wrong sign. However, the situation can be remedied by taking into account the tensorial auxiliary fields.
The equations of motion for them are algebraic:
\begin{equation}
\begin{split}
	T_{(\alpha\beta)} = - 4i \mathcal{F}_{(\alpha\beta)},
	\quad
	&\bar{T}_{(\dot{\alpha}\dot{\beta})}
	=
	4i \bar{\mathcal{F}}_{(\dot{\alpha}\dot{\beta})},
	\\
	T = \bar{T} =0,
	\quad
	P_{\alpha\dot{\alpha}} = \bar{P}_{\alpha\dot{\alpha}} = 0,
	\quad
	&V^{\alpha\dot{\alpha}(ij)} =0,
	\quad
	S^{(ij)} =0\,
\end{split}
\end{equation}
and so these  auxiliary fields can be eliminated  from the action.
As a result of this elimination, we arrive at the on-shell  massless Pauli-Fiertz and Maxwell actions with the correct signs:
\begin{equation}
S^{(bos)}_{lin|on-shell}
=
8 \int d^4x \left(  \Phi^{\alpha\dot{\alpha}\beta\dot{\beta}} \mathcal{G}_{\alpha\dot{\alpha}\beta\dot{\beta}}   +
4 \bar{\mathcal{F}}^{(\dot{\beta}\dot{\rho})}
\bar{\mathcal{F}}_{(\dot{\beta}\dot{\rho})}
+4 \mathcal{F}^{(\beta\rho)}
\mathcal{F}_{(\beta\rho)} \right),
\end{equation}
thereby confirming the status of $C_{\alpha\dot\alpha}$ as the physical ``graviphoton'' field.

In the fermionic sector we are left with the following off-shell action\footnote{Note that, in the process of calculating the component action, many unexpected cancelations occur.}:
\begin{equation}
\begin{split}
	S_{lin}^{(ferm)}
	=
	32i  \int d^4x \Bigr( & \psi^{\beta\alpha\dot{\alpha}i} \bar{\mathcal{G}}_{\beta\alpha\dot{\alpha}i}
	+2 \rho^{\beta i} \partial_\beta^{\dot{\beta}} \bar{\rho}_{\dot{\beta}i}
	+ \frac{1}{16} \rho^{\alpha i} \chi_{\alpha i} + c.c. \Bigr).
\end{split}
\end{equation}
Here $\bar{\mathcal{G}}_{\beta\alpha\dot{\alpha}i}$ is a complex spin $\frac{3}{2}$ gauge field strength (see eqs. (B.3)). The auxiliary fields $\chi^{\alpha i}$ and $\rho^{\alpha i}$ have the algebraic equations of motion
and are equal to zero on shell. So on shell we end up with the free action of the $SU(2)$ doublet of the spin $\frac{3}{2}$ Rarita-Schwinger gauge fields
$\psi^{\beta\alpha\dot{\alpha}i} = \psi^{(\beta\alpha)\dot{\alpha}i} + \epsilon^{\alpha\beta} \psi^{\dot{\alpha} i}$.

Thus we have come to the component off-shell action of the linearized $\mathcal{N}=2$ supergravity and confirmed that on shell it is reduced to the sum of the free spin $2$ action,
the action of the doublet of spins $\frac{3}{2}$ and the spin $1$ action.

\subsection{$R^{mn}R_{mn}$ invariant}

Using the relations of Appendix \ref{app: zero curv solutions}, one can straightforwardly deduce the component contents of the linearized supercurvatures.
The supercurvature   \eqref{eq: SC1} in WZ gauge has the following component expansion:

\begin{equation}
\begin{split}
	\mathcal{F}^{++\alpha\dot{\alpha}} =&
	- 8i \theta^+_\rho \bar{\theta}^+_{\dot{\rho}} \left( \mathcal{R}^{(\alpha\rho)(\dot{\alpha}\dot{\rho})} - \frac{1}{8} \epsilon^{\alpha\rho}\epsilon^{\dot{\alpha}\dot{\rho}} R
	+
	i \epsilon^{(\dot{\alpha}\dot{\rho}} \partial^{\dot{\beta})\rho} P^\alpha_{\dot{\beta}}
	-
	i \epsilon^{(\alpha \rho} \partial^{\beta) \dot{\rho}} \bar{P}^{\dot{\alpha}}_\beta \right)
	\\&+
	4 (\bar{\theta}^+)^2 \left(\partial^{\dot{\alpha}}_\rho T^{(\alpha\rho)} + \partial^{\alpha\dot{\alpha}}T  \right)
	+
	4(\theta^+)^2 \left( \partial^\alpha_{\dot{\rho}} \bar{T}^{(\dot{\alpha}\dot{\rho})} + \partial^{\alpha\dot{\alpha}} \bar{T}\right)
	\\& + \frac{1}{3} V^{\alpha\dot{\alpha}(ij)}u^+_i u^+_j
	+
	\frac{4}{3}i \theta^{+\rho} \bar{\theta}^{+\dot{\rho}} \partial_{\rho\dot{\rho}} V^{\alpha\dot{\alpha}(ij)}u^+_i u^-_j
	-
	\frac{2}{3} (\theta^+)^4 \Box V^{\alpha\dot{\alpha}(ij)}u^-_i u^-_j
	\\&
	-16i \bar{\theta}^+_{\dot{\rho}}\mathcal{G}^{\dot{\rho}\dot{\alpha}\alpha i} u^+_i
	-
	16 i \theta^+_{\rho} \bar{\mathcal{G}}^{\rho\alpha\dot{\alpha}i}u^+_i
	\\&
	+ 32 (\bar{\theta}^+)^2 \theta^{+\rho} \partial_{\rho\dot{\rho}} \mathcal{G}^{\dot{\rho}\dot{\alpha}\alpha i} u^-_i
	-
	32 (\theta^{+})^2 \bar{\theta}^{+\dot{\rho}} \partial_{\rho\dot{\rho}} \bar{\mathcal{G}}^{\rho\alpha\dot{\alpha}i} u^-_i
	\\&
	+
	2i \theta^{+\alpha} \bar{\chi}^{\dot{\alpha}i} u^+_i
	-2i \bar{\theta}^{+\dot{\alpha}} \chi^{\alpha i} u^+_i
	-
	4 (\bar{\theta}^+)^2 \theta^{+\rho} \partial_\rho^{\dot{\alpha}} \chi^{\alpha i } u^-_i
	-
	4 (\theta^+)^2 \bar{\theta}^{+\dot{\rho}} \partial_{\dot{\rho}}^\alpha \bar{\chi}^{\dot{\alpha}i} u^-_i.
\end{split}
\end{equation}
From this expression we conclude that  the analytic superfield $\mathcal{F}^{++\alpha\dot{\alpha}}$
is $\mathcal{N}=2$ generalization of the \textit{linearized Einstein tensor} \eqref{eq: lin Einstein tensor}.

After  integration over Grassmann coordinates and harmonics, the linearized invariant \eqref{eq: inv 1} yields the following component Lagrangian:
\begin{equation}
\begin{split}
	I_1 = -16  \int d^4 x \;
	\Bigr[ & \mathcal{R}^{(\alpha\beta)(\dot{\alpha}\dot{\beta})} \mathcal{R}_{(\alpha\beta)(\dot{\alpha}\dot{\beta})}
	+
	\frac{1}{16} R^2
	\\
	&
	+i \mathcal{R}^{(\alpha\beta)(\dot{\alpha}\dot{\beta})} \partial_{\alpha\dot{\alpha}} P_{\beta\dot{\beta}}
	-
	16i \mathcal{R}^{(\alpha\beta)(\dot{\alpha}\dot{\beta})} \partial_{\alpha\dot{\alpha}} \bar{P}_{\beta\dot{\beta}}
	\\
	& +\frac{3i }{8} \partial_{\beta\dot{\beta}} P^{\beta\dot{\beta}} R
	-
	\frac{3i }{8}\partial_{\beta\dot{\beta}} \bar{P}^{\beta\dot{\beta}} R
	\\&
	-5 P^{\alpha\dot{\alpha}} \Box P_{\alpha\dot{\alpha}}
	-5 \bar{P}^{\alpha\dot{\alpha}} \Box \bar{P}_{\alpha\dot{\alpha}}
	+ 4 P^{\alpha\dot{\beta}} \partial_{\alpha\dot{\alpha}} \partial_{\beta\dot{\beta}} \bar{P}^{\beta\dot{\alpha}}
	\\& - 2 T^{(\alpha\beta)} \partial^{\dot{\alpha}}_\alpha \partial^{\dot{\beta}}_\beta \bar{T}_{(\dot{\alpha}\dot{\beta})}
	+ T \Box T
	+
	\frac{1}{8}\frac{1}{27} V^{\alpha\dot{\alpha}(ij)}\Box V_{\alpha\dot{\alpha}(ij)}
	\\&+ 16 i \mathcal{G}^{\dot{\rho}\dot{\alpha}\alpha i} \partial_{\dot{\rho}}^\rho \bar{\mathcal{G}}_{\rho\alpha\dot{\alpha}i}
	\\&
	+2 i \partial_{\rho\dot{\rho}} \mathcal{G}^{\dot{\rho}\dot{\alpha}\rho i} \bar{\chi}_{\dot{\alpha} i}
	-
	2i \partial_{\rho\dot{\rho}} \mathcal{G}^{\rho\alpha\dot{\rho}i} \chi_{\alpha i}
	-  \frac{i}{2} \bar{\chi}^{\dot{\alpha}i} \partial_{\alpha\dot{\alpha}} \chi^{\alpha}_i
	\Bigr].
\end{split}
\end{equation}
From this component structure we see that $I_1$ is just $\mathcal{N}=2$ superextension of the square of the linearized Ricci curvature.
The auxiliary fields in $I_1$ have non-algebraic equations of motions and so are dynamical, in contract to those in the linearized $\mathcal{N}=2$ Einstein
supergravity\footnote{This property is typical for super-invariants with higher derivatives, see, e.g., \cite{Ivanov:2005qf}.}.

\subsection{$R^2$ invariant}

The component structure of the second  linearized supercurvature \eqref{eq: SC2} is worked out in a similar way:
\begin{equation}
\begin{split}
	\mathcal{F}^{++5} = &
	\frac{i}{2} (\theta^+)^2 \left(R -4i \partial_{\beta\dot{\beta}} P^{\beta\dot{\beta}} \right)- \frac{i}{2} (\bar{\theta}^+) \left( R + i \partial_{\beta\dot{\beta}}\bar{P}^{\beta\dot{\beta}} \right)
	\\&+
	8i \theta^+_\alpha \bar{\theta}^{+\dot{\alpha}}
	\left( \partial^{\alpha\dot{\beta}} \bar{\mathcal{F}}_{(\dot{\alpha}\dot{\beta})} +\frac{i}{2} \partial_{\dot{\alpha}\beta} T^{(\alpha\beta)} \right)
	-8i \bar{\theta}^+_{\dot{\alpha}} \theta^{+\alpha} \left(\partial^{\dot{\alpha}\beta} \mathcal{F}_{(\alpha\beta)}
	- \frac{i}{2} \partial_{\alpha\dot{\beta}} \bar{T}^{(\dot{\alpha}\dot{\beta})}
	\right)
	\\& + \frac{1}{3} S^{(ij)} u^+_i u^+_j
	+
	\frac{4}{3}i \theta^{+\rho}\bar{\theta}^{+\dot{\rho}} \partial_{\rho\dot{\rho}} S^{(ij)} u^+_i u^-_j
	-
	\frac{2}{3} (\theta^+)^4 \Box S^{(ij)} u^-_i u^-_j
	\\& + 4i \theta^{+}_\alpha \mathcal{R}^{\alpha i} u^+_i
	-
	4i \bar{\theta}^+_{\dot{\alpha}} \bar{\mathcal{R}}^{\dot{\alpha}i} u^+_i
	+
	8 (\theta^+)^2 \bar{\theta}^{+\dot{\rho}} \partial_{\dot{\rho}\alpha} \mathcal{R}^{\alpha i }u^-_i
	+
	8 (\bar{\theta}^+)^2 \theta^{+\rho} \partial_{\rho\dot{\alpha}} \bar{\mathcal{R}}^{\dot{\alpha}i} u^-_i
	\\&
	+
	8i \theta^{+\sigma} \partial_\sigma^{\dot{\beta}} \bar{\rho}_{\dot{\beta}}^i u^+_i
	+ 8i \bar{\theta}^{+\dot{\sigma}}\partial_{\dot{\sigma}}^\beta \rho_\beta^i u^+_i
	-
	4 (\bar{\theta}^+)^2 \theta^{+\beta} \Box \rho_\beta^i u^-_i
	+
	4 (\theta^+)^2 \bar{\theta}^{+\dot{\beta}} \Box \bar{\rho}^i_{\dot{\beta}} u^-_i
	\\& - i \theta^{+\alpha} \chi^i_\alpha u^+_i + i \bar{\theta}^{+\dot{\alpha}}\bar{\chi}^i_{\dot{\alpha}} u^+_i
	-
	2 (\theta^+)^2 \bar{\theta}^{+\dot{\rho}} \partial_{\dot{\rho}}^\alpha \chi^i_\alpha u^-_i
	-2 (\bar{\theta}^+)^2 \theta^{+\rho} \partial_\rho^{\dot{\alpha}}\bar{\chi}_{\dot{\alpha}}^i u^-_i.
\end{split}
\end{equation}
Then the component content of the invariant \eqref{eq: inv 2} is given by the expression:
\begin{equation}
\begin{split}
	I_2= \int d^4x \;\Bigr[&\frac{1}{4}R^2 -iR \partial_{\beta\dot{\beta}}\bar{P}^{\beta\dot{\beta}}
	+
	i R \partial_{\beta\dot{\beta}}P^{\beta\dot{\beta}}
	+
	4 (\partial_{\beta\dot{\beta}}P^{\beta\dot{\beta}})(\partial_{\rho\dot{\rho}}\bar{P}^{\rho\dot{\rho}})
	\\
	& +
	\left( 2 \bar{\mathcal{F}}^{(\dot{\alpha}\dot{\beta})} + i\bar{T}^{(\dot{\alpha}\dot{\beta})} \right) \Box \left( 2 \bar{\mathcal{F}}_{(\dot{\alpha}\dot{\beta})} + i\bar{T}_{(\dot{\alpha}\dot{\beta})} \right)
	\\&+
	\left(2\mathcal{F}^{(\alpha\beta)}-i T^{(\alpha\beta)} \right) \Box
	\left(2\mathcal{F}_{(\alpha\beta)}-i T_{(\alpha\beta)} \right)
	\\&
	-8  \left( 2 \bar{\mathcal{F}}^{(\dot{\alpha}\dot{\beta})} + i\bar{T}^{(\dot{\alpha}\dot{\beta})} \right)
	\partial_{\alpha\dot{\alpha}}\partial_{\beta\dot{\beta}}
	\left(2\mathcal{F}^{(\alpha\beta)}-i T^{(\alpha\beta)} \right)
	+
	\dots
	\Bigr].
\end{split}
\end{equation}
This is $\mathcal{N}=2$ generalization of the $R^2$ invariant.

\subsection{Weyl-squared invariant}

Though the component reduction of the linearized $\mathcal{N}=2$ Weyl tensor is technically more involved, it can be performed quite
analogously to the previous supercurvatures. We confine our consideration merely to the lowest term in the supercurvature expansion and to the gravity sector.
As a result, we obtain:
\begin{equation}
\mathcal{W}_{(\alpha\beta)} = 32 \theta^{-(\gamma} \theta^{+\delta)} \mathcal{R}_{(\alpha\beta\gamma\delta)} + \dots,
\end{equation}
\begin{equation}
\mathcal{W} = \theta^- \theta^+ \cdot 0 + \dots.
\end{equation}
From these formulas we observe that $\mathcal{W}$ does not contribute to the pure gravity sector. The terms denoted  by ellipses also contain at least 3 derivatives of the spin 2 field.
So for the invariant $I_3$ defined in \eqref{I3invA} and \eqref{I3invC}  we have the following expression in the gravity sector:
\begin{equation}
I_3 \sim \int d^4x \; \mathcal{R}^{(\alpha\beta\gamma\delta)}\mathcal{R}_{(\alpha\beta\gamma\delta)}.
\end{equation}
The conclusion is that the invariant $I_3$ is an irreducible part of  the linearized $\mathcal{N}=2$ Weyl-squared super-invariant.
It is worth to note that, using $I_3$, its complex conjugate and the invariants $I_1, I_2$, one can also construct
a supersymmetric generalization of the square of Riemann tensor.

\medskip

To summarize, we have the complete  set of the linearized off-shell $\mathcal{N}=2$ supergravity invariants in
terms of analytic prepotentials. In components, they start with the squares of gravity curvatures: $I_1 \sim R^{mn}R_{mn}$, $I_2 \sim R^2$
and $I_3\sim \mathcal{R}^{(\alpha\beta\gamma\delta)}\mathcal{R}_{(\alpha\beta\gamma\delta)}$.

\section{Linearized $\mathcal{N}=2$ supergravity invariants beyond the quadratic order}\label{sec:7}

In harmonic superspace one can construct invariant actions as integrals over  both analytic and full harmonic superspaces:
\begin{equation}
I_{ASS} = \int d\zeta^{(-4)}\, \mathcal{L}^{(+4)},
\qquad
I_{HSS} = \int d^4x d^8\theta du\; \mathcal{L}.
\end{equation}
Using the gauge-invariant supercurvatures $  \mathcal{F}^{++\alpha\dot{\alpha}},     \mathcal{F}^{++5}, \mathcal{W}_{(\alpha\beta)}, \bar{\mathcal{W}}_{(\dot{\alpha}\dot{\beta})}$ defined in \eqref{eq: SC1}, \eqref{eq: SC2}, \eqref{eq: SC3}
and covariant derivatives \eqref{eq: spinor cov}, \eqref{eq:D0}, one can construct a family of such invariant actions.
Being aware of the component structure of supercurvatures, one can generate linearized higher-order $\mathcal{N}=2$ supergravity invariants and deduce their component structure in the gravity sector.

As examples, we present some interesting invariants:

\medskip

$\bullet$ \textit{$R^4$ invariant}
\begin{equation}
\int d^4x d^8\theta du\; \left( \mathcal{D}^{--} \mathcal{F}^{++5} \right)^4 \sim \int d^4x\;  R^4 + \dots\,.
\end{equation}

$\bullet$ \textit{$\Box R \Box R$ invariant}:

\begin{equation}
\int d^4x d^8\theta du\; (\mathcal{D}^-)^2 \mathcal{F}^{++5} (\bar{\mathcal{D}}^-)^2 \mathcal{F}^{++5}
\sim
\int d^4x\;  \Box R \Box R\,.
\end{equation}

$\bullet$ \textit{$R^2 \Box R$ invariant}

\begin{equation}
\int d^4x d^8\theta du\; \left( \mathcal{D}^{--} \mathcal{F}^{++5} \right)^2 (\mathcal{D}^-)^2 \mathcal{F}^{++5} \sim  \int d^4x \; R^2 \Box R\,.
\end{equation}

$\bullet$ \textit{$R^{4+n}$ invariant}
\begin{equation}
\int d^4x d^8\theta du\; \left( \mathcal{D}^{--} \mathcal{F}^{++5} \right)^4 \left[(\mathcal{D}^-)^2 \mathcal{F}^{++5}\right]^n \sim \int d^4x \; R^{4+n}
+
\dots\,.
\end{equation}

$\bullet$ \textit{$G^4$ invariant}
\begin{multline}
\int d^4x d^8\theta du\; \left( \mathcal{D}^{--} \mathcal{F}^{++\alpha_1\dot{\alpha}_1} \right)
\left( \mathcal{D}^{--} \mathcal{F}^{++\alpha_2\dot{\alpha}_2} \right)
\left( \mathcal{D}^{--} \mathcal{F}^{++}_{\alpha_1\dot{\alpha}_2} \right)
\left( \mathcal{D}^{--} \mathcal{F}^{++}_{\alpha_2\dot{\alpha}_1} \right)
\\
\sim\int d^4x\;
\mathcal{G}^{\alpha_1\beta_1 \dot{\alpha}_1 \dot{\beta}_1}
\mathcal{G}^{\alpha_2\beta_2 \dot{\alpha}_2 \dot{\beta}_2}
\mathcal{G}_{\beta_2\alpha_1\dot{\beta}_1 \dot{\alpha}_2}
\mathcal{G}_{\beta_1\alpha_2 \dot{\beta}_2\dot{\alpha}_1} + \dots\,.
\end{multline}

$\bullet$ \textit{Bel-Robinson tensor squared}

Bel-Robinson tensor in the spinor notations is defined as (see, e.g. \cite{Ferrara:1977mv}):
\begin{equation}
\mathcal{R}_{(\alpha\beta\gamma\delta)} \bar{\mathcal{R}}_{(\dot{\alpha}\dot{\beta}\dot{\gamma}\dot{\delta})}.
\end{equation}
Inspecting the component structure of supercurvatures, one immediately observes that Bel-Robinson tensor is contained in the superfield
\begin{equation} \mathcal{W}_{(\alpha\beta)}\bar{\mathcal{W}}_{(\dot{\alpha}\dot{\beta})} \sim \theta^{+\gamma} \theta^{-\delta} \bar{\theta}^{+\dot{\gamma}}\bar{\theta}^{-\dot{\delta}}
\mathcal{R}_{(\alpha\beta\gamma\delta)} \bar{\mathcal{R}}_{(\dot{\alpha}\dot{\beta}\dot{\gamma}\dot{\delta})}.
\end{equation}
Then one can construct $\mathcal{N}=2$ extension of the Bel-Robinson tensor-squared invariant:
\begin{equation}
\int d^4x d^8\theta du \; \mathcal{W}^{(\alpha\beta)} \mathcal{W}_{(\alpha\beta)}\bar{\mathcal{W}}^{(\dot{\alpha}\dot{\beta})} \bar{\mathcal{W}}_{(\dot{\alpha}\dot{\beta})}
\sim
\int d^4x\;
\mathcal{R}^{(\alpha\beta\gamma\delta)}
\mathcal{R}_{(\alpha\beta\gamma\delta)} \bar{\mathcal{R}}^{(\dot{\alpha}\dot{\beta}\dot{\gamma}\dot{\delta})} \bar{\mathcal{R}}_{(\dot{\alpha}\dot{\beta}\dot{\gamma}\dot{\delta})}\,.
\end{equation}

$\bullet$ \textit{$R^3$ invariants and two-loop finiteness of supergravity}

It is known that ${\cal N}=2$ supergravity is finite to two loops. This means the absence of the on-shell cubic supergravity invariants. It is interesting to analyze this issue using our off-shell approach. On dimensional
grounds, $[R^3] = 6$, so such invariants can be constructed as an integral over the full harmonic superspace Lagrangian with the dimension $[\mathcal{L}] = 2$ or as an integral over the analytic superspace with
$[\mathcal{L}^{(+4)}] = 4$. The supercurvatures have dimensions $[\mathcal{F}^{++\alpha\dot{\alpha}}] = 1$,  $[\mathcal{F}^{++5}] = 1$ and $[\mathcal{W}_{(\alpha\beta)}] =1$.

The only cubic invariant which does not vanish on shell is the product of three Riemann tensors.   The product of three Riemann tensors in the spinor notation corresponds
to the invariant
\begin{equation}\label{eq: R^3}
\mathcal{R}_{\alpha\beta}^{\;\;\;\;\gamma\delta}
\mathcal{R}_{\gamma\delta}^{\;\;\;\;\rho\kappa}
\mathcal{R}_{\rho\kappa}^{\;\;\;\;\alpha\beta}.
\end{equation}
The $\mathcal{N}=2$ supersymmetric invariant extending such a product can naturally be constructed  from three $\mathcal{N}=2$ super-Weyl tensors $\mathcal{W}_{(\alpha\beta)}$.
Since $\mathcal{W}_{(\alpha\beta)}$ is not analytic superfield, one needs to integrate over full harmonic superspace or chiral superspace.
Using the component structure, it is easy to prove that  the invariants containing \eqref{eq: R^3} after passing to components,  cannot be constructed in such a way.
Indeed, though a superfield containing \eqref{eq: R^3} can be constructed from the $\mathcal{N}=2$ Weyl tensor,
\begin{equation}
\mathcal{D}^{+\delta} \mathcal{W}_\alpha^{\;\;\beta} \mathcal{W}_\beta^{\;\;\gamma}
\mathcal{D}^-_\delta \mathcal{W}_{\gamma}^{\;\;\alpha}
\sim (\theta^+)^2 (\theta^-)^2   \mathcal{R}_{\alpha\beta}^{\;\;\;\;\gamma\delta}
\mathcal{R}_{\gamma\delta}^{\;\;\;\;\rho\kappa}
\mathcal{R}_{\rho\kappa}^{\;\;\;\;\alpha\beta} + \dots\,,
\end{equation}
this superfield is not chiral and so does not contain the necessary invariant as a coefficient of the highest  $(\theta^+)^4(\theta^-)^4$  monomial.

This reasoning is similar to the  $\mathcal{N}=1$ superspace  arguments about absence of such an invariant in $\mathcal{N}=1$ supergravity \cite{Ferrara:1977mv}.
So we conclude, that it is not possible to $\mathcal{N}=2$ supersymmetrize the invariant \eqref{eq: R^3}.

\section{Conclusions and outlook}    \label{sec:8}
In this paper, using the prepotential formulation of $\mathcal{N}=2$ supergravity in harmonic superspace, we constructed the off-shell linearized $\mathcal{N}=2$ supercurvatures
\begin{equation}\label{eq: SuperCurvatures}
\mathcal{F}^{++\alpha\dot{\alpha}}(\zeta),
\qquad
\mathcal{F}^{++5}(\zeta),
\qquad
\mathcal{W}_{(\alpha\beta)}(x, \theta, u),
\qquad
\bar{\mathcal{W}}_{(\dot{\alpha}\dot{\beta})}(x, \bar{\theta}, u)\,,
\end{equation}
which contain, respectively, the linearized gravity curvatures:
\begin{equation}
\mathcal{G}_{\alpha\beta\dot{\alpha}\dot{\beta}},
\qquad
R,
\qquad
\mathcal{R}_{(\alpha\beta\gamma\delta)},
\qquad
\bar{\mathcal{R}}_{(\dot{\alpha}\dot{\beta}\dot{\gamma}\dot{\delta})}.
\end{equation}
From these supercurvatures we constructed all linearized quadratic invariants and presented some examples
of the linearized invariants beyond the quadratic order.
All the constructed objects are expressed in terms of unconstrained analytical prepotentials, which is the main feature of our approach in contrast to the previous
superspace studies \cite{Butter:2013lta, Kuzenko:2015jxa, Kuzenko:2017uqg, Moura:2002ip, deWit:2010za}. This allows us to systematically derive the component form
of the invariants by solving the harmonic zero-curvature equations in WZ gauge.
It is interesting to note that all the linearized higher-derivative supergravity invariants constructed have an obvious tensor structure
and are manifestly invariant with respect to the gauge transformations. It is the basic difference of these invariants from the action of $\mathcal{N}=2$
Einstein supergravity \eqref{eq: N=2 lin Einstein} which is gauge invariant up to total derivatives and so represents a Chern-Simons-like action.

These results raise a number of interesting and natural issues which we hope to address in the near future.

\medskip
$\bullet$ \textit{Nonlinear invariants in harmonic superspace}
\medskip

The most direct task is to construct nonlinear generalizations of  the supercurvatures and invariants introduced here.
Since there exists no analytical integration measure in this version of $\mathcal{N}=2$ Einstein supergravity \cite{18}, one can expect that
the corresponding nonlinear quadratic invariants can be written in the full harmonic superspace as:
\begin{equation}
\int d^4x d^8\theta du \; E \, H^{--\alpha\dot{\alpha}} (\mathfrak{D}^+)^4 H^{--}_{\alpha\dot{\alpha}},
\qquad
\int d^4x d^8\theta du \; E \, H^{--5} (\mathfrak{D}^+)^4 H^{--5}.
\end{equation}
Checking the invariance of such expressions implies making use of the properties of the integration measure $E$ and integrations by parts.
We hope to achieve some progress toward this goal elsewhere.

It is interesting to establish the connection of such nonlinear invariants with those found in other approaches \cite{Butter:2013lta, Kuzenko:2015jxa,Kuzenko:2017uqg}.
Despite the fact that the  invariants constructed supersymmetrize the same gravitational invariants, it is problematic to compare the results obtained
at the superfield level, since in our approach the invariants are constructed in terms of analytic prepotentials, while the standard approach uses supercurvatures and supertorsions.
This issue is discussed in more detail in the next item. 

\medskip

$\bullet$ \textit{Differential geometry of curved $\mathcal{N}=2$ harmonic superspace}

\medskip

The natural way to construct higher-derivative $\mathcal{N}=2$ supergravity invariants is to use the supertorsions and supercurvatures appearing in the algebra of covariant derivatives \cite{Howe:1981gz}. Such a method was
adopted in \cite{Butter:2013lta, Kuzenko:2015jxa}. The covariant supergeometry of $\mathcal{N}=2$ supergravity in harmonic superspace was considered in \cite{Galperin:1987em, Delamotte:1987yn}. In the harmonic superspace
approach all the basic supergravity constraints are automatically solved in terms of analytic prepotentials. In a separate publication, we shall present such nonlinear invariants constructed out of the covariant torsions and
curvatures, starting from the full harmonic superspace algebra of the covariant derivatives.

\medskip

In this regard, one can also ask which of the linearized supercurvatures \eqref{eq: SuperCurvatures}
have nonlinear extension and how these objects are related to the prepotential superfields solving the $\mathcal{N}=2$ supergravity constraints.
It is well known how the well-defined nonlinear generalizations of invariants $\mathcal{F}^{++5}$ and $\mathcal{W}_{(\alpha\beta)}$ appear
in the covariant superfield formulations of $\mathcal{N}=2$ supergravity \cite{Kuzenko:2017uqg, Grimm:1980kn, Kuzenko:2008ep, Howe:1981gz}.
One would expect\footnote{We thank the referee for bringing this to our attention.}, that a nonlinear extension of the supercurvature
$\mathcal{F}^{++\alpha\dot{\alpha}}$ defined in \eqref{1.2}  can be obtained as a descendant of  the curvature tensor $T^i_\alpha$ of the superspace formulation of
$\mathcal{N}=2$ Poinciar\'e supergravity given in \cite{Castellani:1980cu, Gates:1980ky}. This conjecture is based on the fact that this superspace
formulation of $\mathcal{N}=2$ Einstein supergravity is directly related to the Fradkin-Vasiliev-de Wit-van Holten \cite{Fradkin:1979cw, Fradkin:1979as, deWit:1979xpv}
off-shell $\mathcal{N}=2$ supergravity multiplet which is described just by the unconstrained analytical prepotentials  \eqref{eq: WZ gauge} used through the paper.
Checking this amounts to an additional analysis  at the full nonlinear level and is beyond the scope of the present work.

\medskip
$\bullet$ \textit{$\mathcal{N}=2$ Gauss-Bonnet theorem in harmonic superspace}
\medskip

In the pure gravity, there are two topological invariants quadratic in the Riemann tensor: Pontryagin and Gauss-Bonnet  invariants.
Gauss-Bonnet invariant is given as the integral of Euler density:
\begin{equation}
I_{GB} = \int d^4x \sqrt{g} \left(R^{mnkr} R_{mnkr} - 4 R^{mn}R_{mn}  +R^2 \right).
\end{equation}
Such invariants are topological in the sense that $\delta I_{GB} /\delta g_{mn} = 0$.

$\mathcal{N}=1$ Gauss-Bonnet  invariant was constructed in \cite{Ferrara:1977mv} and its ${\cal N}=2$ cousin in \cite{Butter:2013lta} in the framework of conventional $\mathcal{N}=2$ superspace.
It would be interesting to explicitly construct $\mathcal{N}=2$ generalization of these invariants in harmonic superspace and to explicitly check that such invariants are topological, i.e. satisfy the relation
\begin{equation}
\delta I = \int d^4x d^8\theta du E\; \frac{\delta{I}}{\delta h^{++M}} \delta h^{++M} = 0.
\end{equation}
In  \cite{Butter:2013lta}, the topological nature of $\mathcal{N}=2$ GB-term has been established only at the component level.

\medskip
$\bullet$ \textit{$\mathcal{N}=2$ conformal supergravity invariants}
\medskip

$\mathcal{N}=2$ conformal supergravity formulated in harmonic superspace \cite{Galperin:1987ek} involves an additional analytic prepotential $h^{(+4)}$,
so the structure of the linearized Weyl-squared invariant will be
modified as compared to $\mathcal{N}=2$ Einstein supergravity. It is quite simple to extend the construction of section \ref{sec: QI} to the conformal case.
We shall present this result and its higher-spin generalizations
elsewhere.

The problem of constructing nonlinear $\mathcal{N}=2$ Weyl-squared action in harmonic superspace remains a challenge.

\medskip
$\bullet$ \textit{Other versions of $\mathcal{N}=2$ supergravity}
\medskip

It is well known that various types of $\mathcal{N}=2$ Einstein supergravity can be constructed from the conformal $\mathcal{N}=2$ Weyl multiplet  using superconformal compensators \cite{18, Ivanov:2022vwc, Freedman:2012zz}.
The $\mathcal{N}=2$ Maxwell multiplet must necessarily be present as one of the compensators. The second compensator may be the hypermultiplet, the nonlinear multiplet, the tensor or central-charged multiplets.
These compensators lead to different versions of supergravity.
The ``principal'' version (when one of the compensators is the hypermultiplet) contains an infinite number of off-shell components, other versions contain $\mathbf{40_B} + \mathbf{40_F}$ off-shell degrees of freedom.
In this article, we dealt with the version corresponding to the nonlinear compensating multiplet. It is natural to consider the construction of invariants also in other versions of supergravity.

In this regard, we note the work \cite{Butter:2010sc} in which various linearized versions of $\mathcal{N}=2$ Einstein supergravity were considered.
It is instructive to reproduce such versions within the harmonic superspace framework.
Such formulations can lead to the alternative versions of ${\cal N}=2$ higher spins \cite{Buchbinder:2021ite} and can clarify their relationships with the
off-shell superconformal multiplets of $\mathcal{N}=2$ higher spins recently considered in ref. \cite{Buchbinder:2024pjm}.

It is also of interest to construct the linearized versions of $\mathcal{N}=2$  AdS supergravities, which were investigated in ref. \cite{Butter:2011ym}.
The harmonic formulation of such theories is so far unknown.

\medskip
$\bullet$ \textit{$\mathcal{N}=2$ higher-spin generalizations}
\medskip

Generalizing the invariants constructed to $\mathcal{N}=2$ higher-spin off-shell supermultiplets goes straightforwardly. For example, in terms of the covariant
superfields $G^{\pm\pm\alpha(s-1)\dot{\alpha}(s-1)}$ and
$G^{\pm\pm\alpha(s-2)\dot{\alpha}(s-2)}$ introduced in \cite{Buchbinder:2021ite}, we can immediately derive the following generalizations of the invariants  $I_1$ and $I_2$:
\begin{equation}
\begin{split}
	&I_1^{(s)} =  \int d\zeta^{(-4)} \mathcal{F}^{++\alpha(s-1)\dot{\alpha}(s-1)}\mathcal{F}^{++}_{\alpha(s-1)\dot{\alpha}(s-1)},
	\\
	&
	\mathcal{F}^{++\alpha(s-1)\dot{\alpha}(s-1)} := \left(\mathcal{D}^+ \right)^4 G^{--\alpha(s-1)\dot{\alpha}(s-1)},
\end{split}
\end{equation}
\begin{equation}
\begin{split}
	&I_2^{(s)} = \int d\zeta^{(-4)} \mathcal{F}^{++\alpha(s-2)\dot{\alpha}(s-2)}\mathcal{F}^{++}_{\alpha(s-2)\dot{\alpha}(s-2)},
	\\
	&\mathcal{F}^{++\alpha(s-2)\dot{\alpha}(s-2)} := \left(\mathcal{D}^+ \right)^4 G^{--\alpha(s-2)\dot{\alpha}(s-2)}.
\end{split}
\end{equation}
These invariants are the higher-spin generalizations of the $R^2$ and $R^{m n}R_{m n}$ invariants of $\mathcal{N}=2$ supergravity.
In a similar way, one can generalize other invariants to $\mathcal{N}=2$ higher spins.
In the bosonic case, the higher-spin invariants of this kind were discussed in \cite{Joung:2012qy}.

\medskip
$\bullet$ \textit{Induced actions}
\medskip

It is known that the higher-derivative gravity and higher-spin invariants can be recovered as induced actions by starting from couplings of the higher-spin gauge fields to quantum matter fields
and then integrating  out the matter fields from the generating functional (see, e.g., \cite{BS}). The hypermultiplet $q^+$ is the most general $\mathcal{N}=2$ matter multiplet.
Using cubic interacting vertices of hypermultiplet  with gauge higher-spin superfields \cite{Buchbinder:2022kzl}, one can consider an effective action of the form:
\begin{equation}
\int D q^{+} e^{- \frac{1}{2} \int d\zeta^{(-4)} q^{+a} \left(\mathcal{D}^{++} + \kappa\hat{\mathcal{H}}^{++} \right) q^+_a}
\end{equation}
(see \cite{Buchbinder:2022kzl} for details of the notation). Doing this functional integral must yield some higher-derivative invariants of the higher-spin ${\cal N}=2$ gauge multiplets.
In particular, it would be instructive to reproduce in such a way the invariants introduced at the linearized level in the present paper.

\medskip
$\bullet$ \textit{$6D, \; \mathcal{N}=(1,0)$ supergravity invariants}
\medskip

Various higher-derivative  invariants of $6D, \mathcal{N}=(1,0)$ supergravity have been actively studied in the superfield approach  (see, e.g., \cite{Butter:2016qkx, Butter:2017jqu, Novak:2017wqc, Butter:2018wss}).
$6D$ supergravity in harmonic superspace was constructed by Sokatchev in \cite{Sokatchev:1988aa}. However, the higher-derivative invariants of $\mathcal{N}=(1,0)$ off-shell supergravity
were never tackled in detail by the harmonic superspace methods\footnote{In ref. \cite{BHS} the harmonic approach was applied for analysis of UV divergences of $6D$ supergravities.}.
It is worth noting that the construction of such invariants
in the formulations of ${\cal N}=2$ supergravity without analytic prepotentials requires utilizing new action principles based on the use
of closed covariant super-forms \cite{Butter:2016qkx}. It is of interest to learn whether such invariants admit a reformulation in harmonic superspace. Recall once more that the latter provides
the universal methods of setting up superfield invariants, based upon the analytic gauge prepotentials and the harmonic zero-curvature equations.

\section*{Acknowledgements}

Work of N.Z. was partially supported by the grant 22-1-1-42-2 of
the Foundation for the Advancement of Theoretical Physics and
Mathematics ``BASIS''. The authors are grateful to the anonymous referee for useful and suggestive comments.

\appendix

\section{Linearized gravity invariants in spinor notations}\label{sec:A1}

We use the standard definition of the linearized gravity field $g_{mn} = \eta_{mn} + h_{mn}$, $h:= \eta^{mn} h_{mn}$.
The linearizations of Riemann curvature tensor, Ricci curvature and scalar curvature are given by:
\begin{subequations}\label{eq: linearized curvatures}
\begin{equation}
	R_{krmn} = 2 \partial_{[r} \partial_{[m} h_{n]k]}
	=
	\frac{1}{2} \left( \partial_{m}\partial_{r} h_{nk} - \partial_m \partial_k h_{nr} - \partial_n \partial_r h_{mk}+ \partial_{n} \partial_{k} h_{mr}\right),
\end{equation}
\begin{equation}
	R_{mn} = \frac{1}{2} \left( \partial^r \partial_m h_{nr} + \partial^{r}\partial_n h_{mr}
	-
	\Box h_{mn} - \partial_m \partial_n h \right),
\end{equation}
\begin{equation}
	R = \partial^{m}\partial^n h_{mn} - \Box h.
\end{equation}
\end{subequations}
All curvatures are invariant under the linearized gauge transformations $\delta h_{mn} = \partial_m a_{n} + \partial_n  a_{m}$.

The spin 2 field $\Phi^{\alpha\dot{\alpha}}_{\beta\dot{\beta}}$ from WZ gauge \eqref{eq: WZ gauge} is related to the fierbein as
$\delta_\beta^{\alpha}\delta^{\dot{\alpha}}_{\dot{\beta}}+\Phi^{\alpha\dot{\alpha}}_{\beta\dot{\beta}}
:= \frac{1}{2}\sigma^{\alpha\dot{\alpha}}_m \sigma^a_{\beta\dot{\beta}} e_a^m$.
The fierbein expansion near the flat Minkowski background is given by: $e^m_a = \delta^m_a + \varepsilon_a^m$, so that $h_{mn} = \varepsilon_{mn} + \varepsilon_{nm}$.
After fixing the local Lorentz freedom, in the symmetric gauge $\varepsilon_{mn} = \varepsilon_{(mn)}$ we have :
\begin{equation}
\begin{split}
	h_{mn} = 2 \varepsilon_{mn}
	&=
	\sigma_m^{\alpha\dot{\alpha}} \sigma_n^{\beta\dot{\beta}}
	\left( \Phi_{(\alpha\beta)(\dot{\alpha}\dot{\beta})} + \epsilon_{\alpha\beta} \epsilon_{\dot{\alpha}\dot{\beta}} \Phi \right)
	\\&=
	\sigma_m^{\alpha\dot{\alpha}} \sigma_n^{\beta\dot{\beta}}
	\Phi_{(\alpha\beta)(\dot{\alpha}\dot{\beta})}
	+
	2\eta_{mn} \Phi
	.
\end{split}
\end{equation}
From this relation it follows that  $h= 8\Phi$.

One can rewrite\footnote{Here we use standard notations, see e.g. \cite{18}:
$$\sigma_{[mn]}^{(\alpha\beta)} = \frac{i}{2} \left((\sigma_m)^{\alpha}_{\;\dot{\alpha}} \sigma_n^{\dot{\alpha}\beta} - (\sigma_n)^{\alpha}_{\;\dot{\alpha}} \sigma_m^{\dot{\alpha}\beta}\right),
\qquad
\tilde{\sigma}_{[mn]}^{(\dot{\alpha}\dot{\beta})}
=
\frac{i}{2} \left(\sigma_{m}^{\alpha\dot{\alpha}}
(\sigma_n)_{\alpha}^{\;\,\dot{\beta}}
-
\sigma_{n}^{\alpha\dot{\alpha}}
(\sigma_m)_{\alpha}^{\;\,\dot{\beta}}
\right).$$} the curvatures \eqref{eq: linearized curvatures} in terms of the irreducible spin $2$ fields $\Phi_{(\alpha\beta)(\dot{\alpha}\dot{\beta})}$ and $\Phi$:
\begin{equation}
R_{krmn}
=
\sigma_{[kr]}^{(\alpha\beta)}
\sigma_{[mn]}^{(\gamma\delta)}
\left(-\frac{1}{2}\mathcal{R}_{(\alpha\beta\gamma\delta)} -\frac{1}{48} \epsilon_{\alpha\gamma}\epsilon_{\beta\delta} R \right)
-\frac{1}{4}    \sigma_{[kr]}^{(\alpha\beta)}
\tilde{\sigma}_{[mn]}^{(\dot{\alpha}\dot{\beta})}
\mathcal{R}_{(\alpha\beta)(\dot{\alpha}\dot{\beta})}
+
c.c.,
\end{equation}
\begin{equation}
\begin{split}
	R_{mn}
	=
	\sigma_m^{\alpha\dot{\alpha}} \sigma_n^{\beta\dot{\beta}}
	\Big( \mathcal{R}_{(\alpha\beta)(\dot{\alpha}\dot{\beta})}
	+
	\frac{1}{8}
	\epsilon_{\alpha\beta}\epsilon_{\dot{\alpha}\dot{\beta}} R \Big),
\end{split}
\end{equation}
\begin{equation}\label{eq: R}
R = 4 \partial^{\alpha\dot{\alpha}} \partial^{\beta\dot{\beta}} \Phi_{(\alpha\beta)(\dot{\alpha}\dot{\beta})} - 6 \Box \Phi.
\end{equation}

Here we used the following notations for the irreducible parts of the linearized curvatures:
\begin{equation}\label{eq: conf parts}
\mathcal{R}_{(\alpha\beta\gamma\delta)}
:=
\partial_{(\alpha}^{\dot{\alpha}}\partial_\beta^{\dot{\beta}} \Phi_{\gamma\delta)(\dot{\alpha}\dot{\beta})},
\qquad
\bar{\mathcal{R}}_{(\dot{\alpha}\dot{\beta}\dot{\gamma}\dot{\delta})} := \partial^\alpha_{(\dot{\alpha}} \partial^\beta_{\dot{\beta}} \Phi_{\dot{\gamma}\dot{\delta})(\alpha\beta)},
\end{equation}
\begin{equation}
\mathcal{R}_{(\alpha\beta)(\dot{\alpha}\dot{\beta})}
:=
2\partial_{(\alpha(\dot{\alpha}} \partial^{\rho\dot{\rho}} \Phi_{(\beta)\rho)(\dot{\beta})\dot{\rho})}
-
\frac{1}{2} \Box \Phi_{(\alpha\beta)(\dot{\alpha}\dot{\beta})}
-
2
\partial_{(\alpha(\dot{\alpha}} \partial_{\beta)\dot{\beta})} \Phi.
\end{equation}
Note that the linearized invariants \eqref{eq: conf parts} are constructed from the conformal graviton and so
can be identified with the self-dual and anti-self-dual parts of the linearized Weyl tensor.
For completeness, we give explicit relations for the irreducible decomposition of the Weyl tensor:
\begin{equation}
W_{krmn} := R_{krmn} - \left(g_{k[m} R_{n]r} - g_{r[m} R_{n]k}  \right)
+
\frac{1}{3} R\, g_{k[m} g_{n]r},
\end{equation}
\begin{equation}
W_{krmn}
=
-
\frac{1}{2}
\sigma_{[kr]}^{(\alpha\beta)}
\sigma_{[mn]}^{(\gamma\delta)}
\mathcal{R}_{(\alpha\beta\gamma\delta)}
-
\frac{1}{2}
\tilde{\sigma}_{[kr]}^{(\dot{\alpha}\dot{\beta})}
\tilde{\sigma}_{[mn]}^{(\dot{\gamma}\dot{\delta})}
\bar{\mathcal{R}}_{(\dot{\alpha}\dot{\beta}\dot{\gamma}\dot{\delta})}.
\end{equation}

The linearized Bianchi identities read:
\begin{equation}
\partial^{\beta\dot{\beta}} \mathcal{R}_{(\alpha\beta)(\dot{\alpha}\dot{\beta})} = \frac{1}{8}\partial_{\alpha\dot{\alpha}}R,
\end{equation}
so the linearized Einstein tensor has the form:
\begin{equation}\label{eq: lin Einstein tensor}
\mathcal{G}_{\alpha\beta\dot{\alpha}\dot{\beta}} = \mathcal{R}_{(\alpha\beta)(\dot{\alpha}\dot{\beta})} - \frac{1}{8} \epsilon_{\alpha\beta}\epsilon_{\dot{\alpha}\dot{\beta}}R,
\qquad
\partial^{\beta\dot{\beta}} \mathcal{G}_{\alpha\beta\dot{\alpha}\dot{\beta}} = 0.
\end{equation}
Using the linearized Einstein tensor one can represent the  linearized Einstein-Gilbert action as:
\begin{equation}
S_2 \sim  \int d^4x \; \Phi^{\alpha\beta\dot{\alpha}\dot{\beta}} \mathcal{G}_{\alpha\beta\dot{\alpha}\dot{\beta}}.
\end{equation}

We make use of the above notations for curvatures, when writing solutions of the harmonic zero-curvature equations in the gravity sector, and also
in the component reduction of $\mathcal{N}=2$ invariants.  In such a formulation, the component invariants are explicitly invariant with respect
to the linearized gauge transformations.

\section{Spin $\frac{3}{2}$ linearized curvatures}\label{sec:A2}

In this Appendix, we present  the expressions for the linearized curvatures of the  Rarita-Schwinger spin $\frac{3}{2}$ gauge field.
Such curvatures appear in  solutions of  the zero curvature equations in the fermionic spin $\frac{3}{2}$ sector and in the component  representation
of $\mathcal{N}=2$ supergravity invariants.

In accord with eq. \eqref{eq: spin 3/2 irred parts}, the irreducible parts of the spin $\frac{3}{2}$ Rarita-Schwinger field
undergo the following linearized gauge transformations:
\begin{subequations}
\begin{equation}
	\psi^{\beta\alpha\dot{\alpha}i} =   \psi^{(\beta\alpha)\dot{\alpha}i} + \epsilon^{\beta\alpha} \psi^{\dot{\alpha}i},
	\qquad
	\delta_\lambda \psi^{(\beta\alpha)\dot{\alpha}i} = - \partial^{(\beta\dot{\alpha}}\epsilon^{\alpha)i},
	\quad
	\delta_\lambda \psi^{\dot{\alpha}i} = -\frac{1}{2} \partial^{\dot{\alpha}}_\alpha \epsilon^{\alpha i}
\end{equation}
and, for the complex-conjugated field,
\begin{equation}
	\bar{\psi}^{\dot{\beta}\dot{\alpha}\alpha i} =  \bar{\psi}^{(\dot{\beta}\dot{\alpha})\dot{\alpha}i} + \epsilon^{\dot{\beta}\dot{\alpha}} \bar{\psi}^{\alpha i},
	\qquad
	\delta_\lambda \bar{\psi}^{(\dot{\beta}\dot{\alpha})\alpha i} = - \partial^{(\dot{\beta}\alpha}\bar{\epsilon}^{\dot{\alpha})i},
	\quad
	\delta_\lambda \bar{\psi}^{\alpha i} = -\frac{1}{2} \partial^{\alpha}_{\dot{\alpha}} \epsilon^{\dot{\alpha} i}.
\end{equation}
\end{subequations}

From these irreducible parts one can construct one-derivative irreducible linearized gauge invariants:
\begin{subequations}
\begin{equation}
	\mathcal{R}^i_{(\dot{\alpha}\dot{\beta})\alpha}
	:=
	\partial_{(\dot{\alpha}}^{\beta} \psi^i_{(\alpha\beta)\dot{\beta})}
	+
	\partial_{(\dot{\alpha}\alpha} \psi^i_{\dot{\beta})},
	\qquad
	\mathcal{R}^i_{\alpha} := \partial^{\beta\dot{\beta}}\psi^i_{(\alpha\beta)\dot{\beta}} - 3 \partial_{\alpha}^{\dot{\beta}} \psi^i_{\dot{\beta}},
\end{equation}
\begin{equation}
	\bar{\mathcal{R}}^i_{(\alpha\beta)\dot{\alpha}}
	:=
	\partial_{(\alpha}^{\dot{\beta}} \bar{\psi}^i_{(\dot{\alpha}\dot{\beta})\beta)}
	+
	\partial_{(\alpha\dot{\alpha}} \bar{\psi}^i_{\beta)},
	\qquad
	\bar{\mathcal{R}}^i_{\dot{\alpha}} := \partial^{\beta\dot{\beta}}\bar{\psi}^i_{(\dot{\alpha}\dot{\beta})\beta} - 3 \partial_{\dot{\alpha}}^\beta \bar{\psi}^i_{\beta}.
\end{equation}
\end{subequations}

The Einstein-like spin $\frac{3}{2}$ tensors have the form given below and satisfy the linearized Bianchi identities:
\begin{subequations}\label{eq: BI}
\begin{equation}
	\mathcal{G}^i_{\dot{\alpha}\dot{\beta}\alpha}
	:=
	\mathcal{R}^i_{(\dot{\alpha}\dot{\beta})\alpha}
	+
	\frac{1}{2} \epsilon_{\dot{\alpha}\dot{\beta}} \mathcal{R}^i_\alpha,
	\qquad
	\partial^{\alpha\dot{\alpha}}   \mathcal{G}^i_{\dot{\alpha}\dot{\beta}\alpha} = 0,
\end{equation}
\begin{equation}
	\bar{\mathcal{G}}^i_{\alpha\beta\dot{\alpha}} := \bar{\mathcal{R}}^i_{(\alpha\beta)\dot{\alpha}} +
	\frac{1}{2}\epsilon_{\alpha\beta} \bar{\mathcal{R}}^i_{\dot{\alpha}},
	\qquad
	\partial^{\alpha\dot{\alpha}} \bar{\mathcal{G}}^i_{\alpha\beta\dot{\alpha}}= 0.
\end{equation}
\end{subequations}
Using these linearized tensors, one can construct the gauge-invariant spin $\frac{3}{2}$ action as:
\begin{equation}
S_{\frac{3}{2}} \sim \int d^4x \; i \bar{\psi}^{\dot{\alpha}\dot{\beta}\alpha i} \bar{\mathcal{G}}_{\dot{\alpha}\dot{\beta}\alpha i}.
\end{equation}

\section{Solutions of harmonic zero-curvature equations}\label{app: zero curv solutions}

To perform the component reduction of the supergravity invariants, one needs solutions of four linear harmonic equations \eqref{eq: zero curv eq comp}:
\begin{equation}
\begin{aligned}
	&\mathcal{D}^{++} G^{--\alpha\dot{\alpha}} = \mathcal{D}^{--} G^{++\alpha\dot{\alpha}},
	\\
	&\mathcal{D}^{++} G^{--5} = \mathcal{D}^{--} G^{++5},
	\\
	&\mathcal{D}^{++} G^{--\hat{\alpha}+} = \mathcal{D}^{--} G^{++\hat{\alpha}+}
	\\
	& \mathcal{D}^{++} G^{--\hat{\alpha}+} + G^{--\hat{\alpha}+} = 0.
\end{aligned}
\end{equation}
Because of linearity, one can solve these equations for each field in WZ gauge \eqref{eq: WZ gauge} separately.

Note the important property
\begin{equation}
\delta \left[(\mathcal{D}^+)^4 G^{--\alpha\dot{\alpha}} \right]= 0,
\qquad
\delta \left[ (\mathcal{D}^+)^4 G^{--5} \right] = 0.
\end{equation}
So all $(\theta^-)^4$ components of $G^{--\alpha\dot{\alpha}}$ and $G^{--5}$ are invariant under gauge transformations \eqref{eq: zero-curv G}.
This useful observation makes it possible to verify the correctness of the solutions obtained.
In order to explicitly identify gauge invariant terms in the solutions of zero-curvature equations, we shall use the notation of Appendices \ref{sec:A1} and \ref{sec:A2}.

In this Appendix, we explicitly quote solutions of the zero-curvature equations necessary for the component reduction
of the $\mathcal{N}=2$ supergravity curvatures and invariants defined earlier. We are working with the redefined auxiliary fields $P_{\alpha\dot{\alpha}}$, $\rho_\beta^i$ and $\chi^i_\beta$,
see eqs. \eqref{eq: field redefinition}.

\medskip

\hangindent=0cm \noindent
\textit{\textbf{Gravity sector:}}
\begin{subequations}
\begin{equation}
	G_{(\Phi)}^{++\alpha\dot{\alpha}} = -4i \theta^{+\beta} \bar{\theta}^{+\dot{\beta}} \Phi_{\beta\dot{\beta}}^{\alpha\dot{\alpha}}
	+
	4 (\theta^+)^2 \bar{\theta}^{+\dot{\beta}} \bar{\theta}^{-\dot{\alpha}} B_{\dot{\beta}}^\alpha
	-
	4 (\bar{\theta}^+)^2 \theta^{+\beta} \theta^{-\alpha} B_\beta^{\dot{\alpha}},
\end{equation}
\begin{equation}
	\begin{split}
		G^{--\alpha\dot{\alpha}}_{(\Phi)}
		=
		&-4i \theta^{-\beta} \bar{\theta}^{-\dot{\beta}} \Phi^{\alpha\dot{\alpha}}_{\beta\dot{\beta}}
		+
		8 (\theta^-)^2 \bar{\theta}^{-\dot{\rho}} \bar{\theta}^{+\dot{\beta}} \left( \partial^\beta_{(\dot{\rho}} \Phi^{\alpha\dot{\alpha}}_{\beta\dot{\beta})}
		- \frac{1}{2} \delta^{\dot{\alpha}}_{\dot{\rho}} B^\alpha_{\dot{\beta}} \right)
		\\&
		-
		8 (\bar{\theta}^-)^2 \theta^{-\rho} \theta^{+\beta} \left( \partial_{(\rho}^{\dot{\beta}} \Phi^{\alpha\dot{\alpha}}_{\dot{\beta}\beta)} - \frac{1}{2} \delta^\alpha_\rho  B_\beta^{\dot{\alpha}}\right)
		\\&-
		8i (\theta^-)^4 \theta^+_\rho \bar{\theta}^+_{\dot{\rho}}
		\left( \mathcal{R}^{(\alpha\rho)(\dot{\alpha}\dot{\rho})} - \frac{1}{8} \epsilon^{\alpha\rho} \epsilon^{\dot{\alpha}\dot{\rho}} R \right),
	\end{split}
\end{equation}
\medskip
\begin{equation}
	G_{(\Phi)}^{++5} = 2 (\theta^+)^2 \bar{\theta}^{+\dot{\beta}} \theta^{-\beta} B_{\beta\dot{\beta}}
	+
	2 (\bar{\theta}^+)^2 \theta^{+\beta} \bar{\theta}^{-\dot{\beta}} B_{\beta\dot{\beta}},
\end{equation}
\begin{equation}
	\begin{split}
		G_{(\Phi)}^{--5} = &- 2 (\theta^-)^2 \bar{\theta}^{-\dot{\beta}} \theta^{+\beta} B_{\beta \dot{\beta}}
		-
		\frac{i}{2} (\theta^-)^4 (\theta^+)^2 R
		\\
		&
		- 2 (\bar{\theta}^-)^2 \theta^{-\beta} \bar{\theta}^{+\dot{\beta}} B_{\beta\dot{\beta}}
		+
		\frac{i}{2} (\theta^-)^4 (\bar{\theta}^+)^2 R,
	\end{split}
\end{equation}
\medskip
\begin{equation}
	G^{++\alpha+}_{(\Phi)}
	=
	-i
	(\theta^+)^2 \bar{\theta}^+_{\dot{\beta}} B^{\alpha\dot{\beta}},
\end{equation}
\begin{equation}
	\begin{split}
		G^{--\alpha+}_{(\Phi)}
		=&
		-
		\frac{i}{2} (\theta^-)^2 \bar{\theta}^+_{\dot{\beta}}B^{\alpha\dot{\beta}}
		-i
		(\theta^+\theta^-) \bar{\theta}^-_{\dot{\beta}} B^{\alpha\dot{\beta}}
		\\&
		+(\theta^+)^2 (\bar{\theta}^-)^2 \theta^{-\rho} \partial_{\rho\dot{\beta}} B^{\alpha\dot{\beta}}
		-
		(\theta^-)^2 (\bar{\theta}^+ \bar{\theta}^-) \theta^{+\rho}\partial_{\rho\dot{\beta}}B^{\alpha\dot{\beta}}
		\\& +2 (\theta^-)^2 \theta^{+\rho}\bar{\theta}^{+(\dot{\rho}} \theta^{-\dot{\beta})} \partial_{\rho\dot{\rho}} B^\alpha_{\dot{\beta}}
		-
		\frac{i}{2} (\theta^-)^4 (\theta^+)^2 \bar{\theta}^{+\dot{\beta}} \Box B^\alpha_{\dot{\beta}},
	\end{split}
\end{equation}
\begin{equation}
	G^{--\alpha-}_{(\Phi)}
	=
	-\frac{i}{2}(\theta^-)^2 \bar{\theta}^-_{\dot{\beta}} B^{\alpha\dot{\beta}}
	+
	(\theta^-)^4 \theta^{+\beta} \partial_{\beta\dot{\beta}} B^{\alpha\dot{\beta}}.
\end{equation}
\end{subequations}
\textbf{\textit{Maxwell sector:}}
\begin{subequations}
\begin{equation}
	G^{++5}_{(C)} = -4i \theta^{+\beta}\bar{\theta}^{+\dot{\beta}} C_{\beta\dot{\beta}}\,,
\end{equation}
\begin{equation}
	\begin{split}
		G^{--5}_{(C)} =& - 4i \theta^{-\beta} \bar{\theta}^{-\dot{\beta}} C_{\beta\dot{\beta}}
		\\
		&+
		8 (\theta^-)^2 \bar{\theta}^{-(\dot{\rho}} \bar{\theta}^{+\dot{\beta})} \bar{\mathcal{F}}_{(\dot{\rho}\dot{\beta})}
		-
		8 (\bar{\theta}^-)^2 \theta^{-(\rho} \theta^{+\beta)} \mathcal{F}_{(\rho\beta)}
		\\
		& - 8i (\theta^-)^4 \theta^{+\rho} \bar{\theta}^{+\dot{\rho}} \left( \partial_{\rho}^{\dot{\beta}} \bar{\mathcal{F}}_{(\dot{\rho}\dot{\beta})}+ \partial_{\dot{\rho}}^\beta \mathcal{F}_{(\rho\beta)} \right).
	\end{split}
\end{equation}
\end{subequations}
\textbf{\textit{Auxiliary field $T^{(\alpha\beta)}$:}}
\begin{subequations}
\begin{equation}
	G_{T^{(\alpha\beta)}}^{++\alpha\dot{\alpha}} = 4i (\bar{\theta}^+)^2 \theta^+_\beta \bar{\theta}^{-\dot{\alpha}} T^{(\alpha\beta)}
	-
	4i (\theta^+)^2 \bar{\theta}^+_{\dot{\beta}}\theta^{-\alpha} \bar{T}^{(\dot{\alpha}\dot{\beta})},
\end{equation}
\begin{equation}
	\begin{split}
		G_{T^{(\alpha\beta)}}^{--\alpha\dot{\alpha}} =&
		-4i (\bar{\theta}^-)^2 \theta^-_\beta \bar{\theta}^{+\dot{\alpha}}  T^{(\alpha\beta)}
		+
		4 (\theta^-)^4 (\bar{\theta}^+)^2 \partial^{\dot{\alpha}}_\rho T^{(\alpha\rho)}
		\\&
		+
		4i (\theta^-)^2 \bar{\theta}^-_{\dot{\beta}} \theta^{+\alpha} \bar{T}^{(\dot{\alpha}\dot{\beta})}
		+
		4(\theta^-)^4 (\theta^+)^2 \partial^\alpha_{\dot{\rho}} \bar{T}^{(\dot{\alpha}\dot{\rho})},
	\end{split}
\end{equation}

\begin{equation}
	G_{T^{(\alpha\beta)}}^{++5}
	=
	-2i (\bar{\theta}^+)^2 \theta^+_{\alpha}\theta^-_\beta T^{(\alpha\beta)}
	-
	2i (\theta^+)^2 \bar{\theta}^+_{\dot{\alpha}} \bar{\theta}^-_{\dot{\beta}} \bar{T}^{(\dot{\alpha}\dot{\beta})},
\end{equation}
\begin{equation}
	\begin{split}
		G_{T^{(\alpha\beta)}}^{--5}
		=&
		-2i (\bar{\theta}^-)^2 \theta^+_\alpha \theta^-_\beta T^{(\alpha\beta)}
		-
		4 (\theta^-)^4 \theta^+_\alpha \bar{\theta}^{+\dot{\sigma}} \partial_{\beta\dot{\sigma}} T^{(\alpha\beta)}
		\\
		& -2i  (\theta^-)^2 \bar{\theta}^+_{\dot{\alpha}}
		\bar{\theta}^-_{\dot{\beta}} \bar{T}^{(\dot{\alpha}\dot{\beta})}
		-4
		(\theta^-)^4
		\bar{\theta}^{+}_{\dot{\alpha}} \theta^{+\sigma} \partial_{\sigma\dot{\beta}} \bar{T}^{(\dot{\alpha}\dot{\beta})},
	\end{split}
\end{equation}

\medskip

\begin{equation}
	G_{T^{(\alpha\beta)}}^{++\alpha+} = - (\bar{\theta}^+)^2 \theta^+_\beta T^{(\alpha\beta)},
\end{equation}
\begin{equation}
	\begin{split}
		G_{T^{(\alpha\beta)}}^{--\alpha+} = &- \frac{1}{2} (\bar{\theta}^-)^2 \theta^+_\beta T^{(\alpha\beta)}
		-
		(\bar{\theta}^+\bar{\theta}^-) \theta^-_\beta T^{(\alpha\beta)}
		\\&+
		i (\bar{\theta}^-)^2 (\theta^+ \theta^-) \bar{\theta}^{+\dot{\rho}} \partial_{\dot{\rho}\beta} T^{(\alpha\beta)}
		-
		2i (\bar{\theta}^-)^2 \theta^{+(\beta} \theta^{-\rho)} \bar{\theta}^{+\dot{\rho}} \partial_{\rho\dot{\rho}} T^{(\alpha}_{\beta)}
		\\& - i (\bar{\theta}^+)^2 (\theta^-)^2 \bar{\theta}^{-\dot{\rho}} \partial_{\dot{\rho}\beta} T^{(\alpha\beta)}
		+
		\frac{1}{2}(\theta^-)^4 (\bar{\theta}^+)^2 \theta^+_\beta \Box T^{(\alpha\beta)},
	\end{split}
\end{equation}
\begin{equation}
	G_{T^{(\alpha\beta)}}^{--\alpha-}
	=
	\frac{1}{2} (\bar{\theta}^-)^2 \theta^-_\beta T^{(\alpha\beta)}
	-
	i (\theta^-)^4 \bar{\theta}^{+\dot{\rho}} \partial_{\dot{\rho}\beta}T^{(\alpha\beta)}.
\end{equation}
\end{subequations}
\textit{\textbf{Auxiliary field $T$:}}
\begin{subequations}
\begin{equation}
	G^{++\alpha\dot{\alpha}}_{(T)} =
	4i  (\bar{\theta}^+)^2 \theta^{+\alpha} \bar{\theta}^{-\dot{\alpha}} T
	-
	4i (\theta^+)^2 \bar{\theta}^{+\dot{\alpha}} \theta^{-\alpha} \bar{T}\,,
\end{equation}
\begin{equation}
	\begin{split}
		G^{--\alpha\dot{\alpha}}_{(T)} = &
		-4i (\bar{\theta}^-)^2 \theta^{-\alpha}\bar{\theta}^{+\dot{\alpha}} T
		+
		4 (\theta^-)^4 (\bar{\theta}^+)^2 \partial^{\alpha\dot{\alpha}} T
		\\& +4i (\theta^-)^2 \bar{\theta}^{-\dot{\alpha}} \theta^{+\alpha} \bar{T}
		+
		4 (\theta^-)^4 (\theta^+)^2 \partial^{\alpha\dot{\alpha}} \bar{T},
	\end{split}
\end{equation}
\medskip
\begin{equation}
	G^{++5}_{(T)} =  -2i (\bar{\theta}^+)^2 (\theta^+ \theta^-) T
	+
	2i (\theta^+)^2 (\bar{\theta}^+ \bar{\theta}^-) \bar{T},
\end{equation}\begin{equation}
	G^{--5}_{(T)} =  -2i (\theta^-)^2 (\bar{\theta}^+ \bar{\theta}^-) T
	+
	2i (\bar{\theta}^-)^2 (\theta^+ \theta^-) \bar{T},
\end{equation}
\medskip
\begin{equation}
	G^{++\alpha+}_{(T)} = - (\bar{\theta}^+)^2 \theta^{+\alpha} T,
\end{equation}
\begin{equation}
	\begin{split}
		G^{--\alpha+}_{(T)} =&
		-
		\frac{1}{2} (\bar{\theta}^-)^2 \theta^{+\alpha} T - (\bar{\theta}^+ \bar{\theta}^-) \theta^{-\alpha} T
		\\& - i (\bar{\theta}^+)^2 (\theta^-)^2 \bar{\theta}^{-\dot{\rho}} \partial_{\dot{\rho}}^\alpha T
		+
		i
		(\bar{\theta}^-)^2
		(\theta^+\theta^-)  \bar{\theta}^{+\dot{\rho}} \partial_{\dot{\rho}}^\alpha T
		\\& - 2i (\bar{\theta}^-)^2 \theta^{+(\alpha} \theta^{-\rho)} \bar{\theta}^{+\dot{\rho}} \partial_{\rho\dot{\rho}} T
		+
		\frac{1}{4} (\theta^-)^4 (\bar{\theta}^+)^2 \theta^{-\alpha} \Box T,
	\end{split}
\end{equation}
\begin{equation}
	G^{--\alpha-}_{(T)} =  \frac{1}{2} (\bar{\theta}^-)^2 \theta^{-\alpha} T
	-
	\frac{i}{2}(\theta^-)^4 \bar{\theta}^{+\dot{\rho}} \partial_{\dot{\rho}}^\alpha T.
\end{equation}

\end{subequations}
\textit{\textbf{Auxiliary field $P^{\alpha\dot{\alpha}}$:}}
\begin{subequations}
\begin{equation}
	G^{++\alpha\dot{\alpha}}_{(P)} =
	4i (\theta^+)^2 \bar{\theta}^+_{\dot{\beta}} \bar{\theta}^{-\dot{\alpha}} P^{\alpha\dot{\beta}}
	+
	4i (\bar{\theta}^+)^2 \theta^+_\beta \theta^{-\alpha} \bar{P}^{\beta\dot{\alpha}},
\end{equation}
\begin{equation}
	\begin{split}
		G^{--\alpha\dot{\alpha}}_{(P)} =& 2i (\bar{\theta}^-)^2 (\theta^+ \theta^-) P^{\alpha\dot{\alpha}}
		+
		4i (\theta^-)^2 \bar{\theta}^+_{(\dot{\beta}} \bar{\theta}^{-\dot{\alpha})} P^{\alpha\dot{\beta}} - 8 (\theta^-)^4 \theta^{+\rho} \bar{\theta}^+_{(\dot{\beta}} \partial^{\dot{\alpha})}_\rho P^{\alpha\dot{\beta}}
		\\&- 2i (\theta^-)^2 (\bar{\theta}^+ \bar{\theta}^-) \bar{P}^{\alpha\dot{\alpha}}
		+
		4i (\bar{\theta}^-)^2 \theta^{+}_{(\beta} \theta^{-\alpha)} \bar{P}^{\beta\dot{\alpha}}
		-
		8 (\theta^-)^4 \bar{\theta}^{+\dot{\rho}} \theta^+_{(\beta} \partial^{\alpha)}_{\dot{\rho}} \bar{P}^{\beta\dot{\alpha}},
	\end{split}
\end{equation}
\begin{equation}
	G^{++5}_{(P)} =
	2i (\theta^+)^2 \theta^-_\beta \bar{\theta}^+_{\dot{\beta}} P^{\beta\dot{\beta}}
	+
	2i (\bar{\theta}^+)^2 \theta^+_\beta \bar{\theta}^-_{\dot{\beta}} \bar{P}^{\beta\dot{\beta}},
\end{equation}
\begin{equation}
	\begin{split}
		G^{--5}_{(P)} =& -2i (\theta^-)^2 \theta^+_\beta \bar{\theta}^-_{\dot{\beta}} P^{\beta\dot{\beta}}
		+
		2 (\theta^-)^4 (\theta^+)^2 \partial_{\beta\dot{\beta}} P^{\beta\dot{\beta}}
		\\&
		-2i (\bar{\theta}^-)^2 \theta^-_\beta \bar{\theta}^+_{\dot{\beta}} \bar{P}^{\beta\dot{\beta}}
		+
		2 (\theta^-)^4 (\bar{\theta}^+)^2 \partial_{\beta\dot{\beta}} \bar{P}^{\beta\dot{\beta}},
	\end{split}
\end{equation}
\medskip
\begin{equation}
	G^{++\alpha+}_{(P)}
	=
	-
	(\theta^+)^2 \bar{\theta}^+_{\dot{\beta}} P^{\alpha\dot{\beta}},
\end{equation}
\begin{equation}
	\begin{split}
		G^{--\alpha+}_{(P)}
		=&
		-
		\frac{1}{2} (\theta^-)^2 \bar{\theta}^+_{\dot{\beta}}P^{\alpha\dot{\beta}}
		-
		(\theta^+\theta^-) \bar{\theta}^-_{\dot{\beta}} P^{\alpha\dot{\beta}}
		\\&
		-
		i (\theta^+)^2 (\bar{\theta}^-)^2 \theta^{-\rho} \partial_{\rho\dot{\beta}} P^{\alpha\dot{\beta}}
		+
		i (\theta^-)^2 (\bar{\theta}^+ \bar{\theta}^-) \theta^{+\rho}\partial_{\rho\dot{\beta}}P^{\alpha\dot{\beta}}
		\\& -2i (\theta^-)^2 \theta^{+\rho}\bar{\theta}^{+(\dot{\rho}} \theta^{-\dot{\beta})} \partial_{\rho\dot{\rho}} P^\alpha_{\dot{\beta}}
		-
		\frac{1}{2} (\theta^-)^4 (\theta^+)^2 \bar{\theta}^{+\dot{\beta}} \Box P^\alpha_{\dot{\beta}},
	\end{split}
\end{equation}
\begin{equation}
	G^{--\alpha-}_{(P)}
	=
	-\frac{1}{2}(\theta^-)^2 \bar{\theta}^-_{\dot{\beta}} P^{\alpha\dot{\beta}}
	- i
	(\theta^-)^4 \theta^{+\beta} \partial_{\beta\dot{\beta}} P^{\alpha\dot{\beta}}.
\end{equation}
\end{subequations}
\textit{\textbf{Auxiliary field $V^{\alpha\dot{\alpha}(ij)}$:}}
\begin{subequations}
\begin{equation}
	G^{++\alpha\dot{\alpha}}_{(V)}
	=
	(\theta^+)^4 V^{\alpha\dot{\alpha}(ij)} u^-_i u^-_j\,,
\end{equation}
{ \begin{equation}
		\begin{split}
			G^{--\alpha\dot{\alpha}}_{(V)} =&
			\frac{1}{3} (\theta^-)^2 (\bar{\theta}^+)^2 V^{\alpha\dot{\alpha}(ij)} u^-_i u^-_j
			+
			\frac{1}{3} (\bar{\theta}^-)^2 (\theta^+)^2 V^{\alpha\dot{\alpha}(ij)}u^-_i u^-_j
			+
			\frac{4}{3} (\theta^+ \theta^-) (\bar{\theta}^+ \bar{\theta}^-) V^{\alpha\dot{\alpha}(ij)}u^-_i u^-_j
			\\&-\frac{2}{3} (\theta^-)^2 (\bar{\theta}^+ \bar{\theta}^-) V^{\alpha\dot{\alpha}(ij)}u^+_iu^-_j
			-
			\frac{2}{3} (\bar{\theta}^-)^2 (\theta^+\theta^-) V^{\alpha\dot{\alpha}(ij)} u^+_i u^-_j
			\\&+
			\frac{4}{3}i (\theta^+)^2 (\bar{\theta}^-)^2 \theta^{-\rho}\bar{\theta}^{+\dot{\rho}} \partial_{\rho\dot{\rho}} V^{\alpha\dot{\alpha}(ij)} u^-_i u^-_j
			+
			\frac{4}{3}i (\bar{\theta}^+)^2 (\theta^-)^2 \theta^{+\rho} \bar{\theta}^{-\dot{\rho}} \partial_{\rho\dot{\rho}} V^{\alpha\dot{\alpha}(ij)} u^-_i u^-_j
			\\&
			+
			\frac{1}{3} (\theta^-)^4 V^{\alpha\dot{\alpha}(ij)}u^+_i u^+_j
			+
			\frac{4}{3}i (\theta^-)^4 \theta^{+\rho} \bar{\theta}^{+\dot{\rho}} \partial_{\rho\dot{\rho}} V^{\alpha\dot{\alpha}(ij)} u^+_i u^-_j
			-
			\frac{2}{3} (\theta^-)^4 (\theta^+)^4 \Box V^{\alpha\dot{\alpha}(ij)} u^-_i u^-_j.
		\end{split}
\end{equation}  }
\end{subequations}
\textit{\textbf{Auxiliary field $S^{(ij)}$:}}
\begin{subequations}
\begin{equation}
	G^{++5}
	_{(S)} = (\theta^+)^4 S^{(ij)}u^-_i u^-_j,
\end{equation}
\begin{equation}
	\begin{split}
		G^{--5}
		_{(S)} =&
		\frac{1}{3} (\theta^-)^2 (\bar{\theta}^+)^2 S^{(ij)} u^-_i u^-_j
		+
		\frac{1}{3} (\bar{\theta}^-)^2 (\theta^+)^2 S^{(ij)}u^-_i u^-_j
		+
		\frac{4}{3} (\theta^+ \theta^-) (\bar{\theta}^+ \bar{\theta}^-) S^{(ij)}u^-_i u^-_j
		\\&-\frac{2}{3} (\theta^-)^2 (\bar{\theta}^+ \bar{\theta}^-) S^{(ij)} u^+_iu^-_j
		-
		\frac{2}{3} (\bar{\theta}^-)^2 (\theta^+\theta^-) S^{(ij)}u^+_i u^-_j
		\\&+
		\frac{4}{3}i (\theta^+)^2 (\bar{\theta}^-)^2 \theta^{-\rho}\bar{\theta}^{+\dot{\rho}} \partial_{\rho\dot{\rho}} S^{(ij)} u^-_i u^-_j
		+
		\frac{4}{3}i (\bar{\theta}^+)^2 (\theta^-)^2 \theta^{+\rho} \bar{\theta}^{-\dot{\rho}} \partial_{\rho\dot{\rho}} S^{(ij)}u^-_i u^-_j
		\\&
		+
		\frac{1}{3} (\theta^-)^4 S^{ij} u^+_i u^+_j
		+
		\frac{4}{3}i (\theta^-)^4 \theta^{+\rho} \bar{\theta}^{+\dot{\rho}} \partial_{\rho\dot{\rho}} S^{(ij)}u^+_i u^-_j
		- \frac{2}{3} (\theta^-)^4 (\theta^+)^4 \Box S^{(ij)} u^-_i u^-_j.
	\end{split}
\end{equation}
\end{subequations}
\textit{\textbf{Rarita-Schwinger field $\psi_{}^{\alpha\beta\dot{\alpha}i} =\psi_{}^{(\alpha\beta)\dot{\alpha}i} + \epsilon^{\alpha\beta} \psi^{\dot{\alpha} i}$:}}

\begin{subequations}
\begin{equation}
	\begin{split}
		G_{(\psi)}^{++\alpha\dot{\alpha}} =& 16 (\bar{\theta}^+)^2 \theta^{+\beta} \psi_\beta^{\alpha\dot{\alpha}i} u^-_i
		-
		16(\theta^+)^2 \bar{\theta}^{+\dot{\beta}} \bar{\psi}_{\dot{\beta}}^{\alpha\dot{\alpha}i}u^-_i
		\\&
		- 16i (\theta^+)^4 \bar{\theta}^{-\dot{\beta}} \partial_{\dot{\beta}}^\beta \psi^{\alpha\dot{\alpha} i}_\beta u^-_i
		-
		16i (\theta^+)^4 \theta^{-\beta} \partial_\beta^{\dot{\beta}}\bar{\psi}^{\alpha\dot{\alpha}i}_{\dot{\beta}}u^-_i,
	\end{split}
\end{equation}
\begin{equation}
	\begin{split}
		G_{(\psi)}^{--\alpha\dot{\alpha}}
		=
		&\quad
		16(\bar{\theta}^+\bar{\theta}^-) \theta^{-\beta} \psi_\beta^{\alpha\dot{\alpha}i}  u^-_i
		+
		8 (\bar{\theta}^-)^2 \theta^{+\beta} \psi_\beta^{\alpha\dot{\alpha}i} u^-_i
		-
		8 (\bar{\theta}^-)^2 \theta^{-\beta} \psi_\beta^{\alpha\dot{\alpha}i} u^+_i
		\\&
		-16i (\bar{\theta}^-)^2 (\theta^+ \theta^-) \bar{\theta}^{+\dot{\sigma}} \left(\partial_{\dot{\sigma}}^\beta \psi_\beta^{\alpha\dot{\alpha}i}
		- \delta^{\dot{\alpha}}_{\dot{\sigma}} \partial^\beta_{\dot{\beta}} \psi_\beta^{\alpha\dot{\beta}i} \right)u^-_i
		\\&
		+
		8i (\bar{\theta}^+)^2 (\theta^-)^2 \bar{\theta}^{-\dot{\sigma}} \left(2\partial_{\dot{\sigma}}^\beta \psi_\beta^{\alpha\dot{\alpha}i}
		-
		\delta^{\dot{\alpha}}_{\dot{\sigma}}
		\partial^\beta_{\dot{\beta}} \psi_\beta^{\alpha\dot{\beta}i} \right)u^-_i
		\\&
		-32i (\bar{\theta}^-)^2 \theta^{-(\beta} \theta^{+\sigma)} \bar{\theta}^{+\dot{\sigma}} \partial_{\sigma\dot{\sigma}} \psi_\beta^{\alpha\dot{\alpha}i} u^-_i
		\\&-
		16i (\theta^-)^4 \bar{\theta}^{+}_{\dot{\sigma}}
		\mathcal{G}^{\dot{\sigma}\dot{\alpha}\alpha i} u^+_i
		+32(\theta^-)^4 (\bar{\theta}^+)^2 \theta^{+\rho} \partial_{\rho\dot{\rho}} \mathcal{G}^{\dot{\rho}\dot{\alpha}\alpha i}u^-_i
		\\&
		-16 (\theta^+\theta^-) \bar{\theta}^{-\dot{\beta}} \bar{\psi}^{\alpha\dot{\alpha}i}_{\dot{\beta}} u^-_i
		-
		8 (\theta^-)^2 \bar{\theta}^{+\dot{\beta}} \bar{\psi}^{\alpha\dot{\alpha}i}_{\dot{\beta}} u^-_i
		+
		8 (\theta^-)^2 \bar{\theta}^{-\dot{\beta}} \bar{\psi}^{\alpha\dot{\alpha}i}_{\dot{\beta}} u^+_i
		\\& -16i (\theta^-)^2(\bar{\theta}^+ \bar{\theta}^-) \theta^{+\sigma} \left(  \partial_\sigma^{\dot{\beta}} \bar{\psi}^{\alpha\dot{\alpha}i}_{\dot{\beta}} -  \delta_\sigma^\alpha \partial_\beta^{\dot{\beta}} \bar{\psi}_{\dot{\beta}}^{\beta\dot{\alpha}i}  \right)u^-_i
		\\&
		+
		8i (\theta^+)^2 (\bar{\theta}^-)^2 \theta^{-\sigma}
		\left( 2\partial_\sigma^{\dot{\beta}} \bar{\psi}^{\alpha\dot{\alpha}i}_{\dot{\beta}} - \delta^\alpha_\sigma \partial^{\dot{\beta}}_\beta \bar{\psi}^{\beta\dot{\alpha}i}_{\dot{\beta}} \right) u^-_i
		\\&
		+
		32 i (\theta^-)^2 \bar{\theta}^{-(\dot{\beta}} \bar{\theta}^{+\dot{\sigma})} \theta^{+\sigma} \partial_{\sigma\dot{\sigma}} \bar{\psi}^{\alpha\dot{\alpha}i}_{\dot{\beta}} u^-_i
		\\&
		-16i (\theta^-)^4 \theta^{+}_{\sigma} \bar{\mathcal{G}}^{\sigma\alpha\dot{\alpha} i} u^+_i
		- 32 (\theta^-)^4 (\theta^+)^2 \bar{\theta}^{+\dot{\rho}} \partial_{\rho\dot{\rho}} \bar{\mathcal{G}}^{\rho\alpha\dot{\alpha}i}u^-_i,
	\end{split}
\end{equation}

\medskip

\begin{equation}
	\begin{split}
		G^{++5}_{(\psi)} = &-16 (\theta^+)^2\bar{\theta}^{+\dot{\beta}} \psi^i_{\dot{\beta}}u^-_i +  8i (\theta^+)^4 \theta^{-}_{\alpha} \partial^\beta_{\dot{\beta}}\psi_\beta^{\alpha\dot{\beta}i} u^-_i
		\\&
		+16 (\bar{\theta}^+)^2 \theta^{+\beta} \bar{\psi}^i_\beta u^-_i
		-8i (\theta^+)^4 \bar{\theta}^{-}_{\dot{\alpha}} \partial_\beta^{\dot{\beta}}\bar{\psi}_{\dot{\beta}}^{\beta\dot{\alpha}i} u^-_i,
	\end{split}
\end{equation}
\begin{equation}
	\begin{split}
		G^{--5}_{(\psi)} =
		&-16 (\theta^+\theta^-) \bar{\theta}^{-\dot{\beta}} \psi^i_{\dot{\beta}}u^-_i
		-
		8
		(\theta^-)^2 \bar{\theta}^{+\dot{\beta}}\psi^i_{\dot{\beta}} u^-_i
		+
		8 (\theta^-)^2 \bar{\theta}^{-\dot{\beta}} \psi_{\dot{\beta}}^i u^+_i
		\\
		&
		- 8i (\theta^-)^2 (\bar{\theta}^+ \bar{\theta}^-)  \theta^{+}_{\alpha} \mathcal{R}^{\alpha i} u^-_i
		-4
		i (\theta^+)^2 (\bar{\theta}^-)^2 \theta^{-}_{\alpha} (\mathcal{R}^{\alpha i} + 2 \partial^{\alpha\dot{\beta}}\psi^-_{\dot{\beta}}) u^-_i
		\\&+ 4 i (\theta^-)^4 \theta^{+}_{\alpha}\mathcal{R}^{\alpha i} u^+_i
		+
		8 (\theta^-)^4 (\theta^+)^2 \bar{\theta}^{+\dot{\rho}} \partial_{\dot{\rho}\alpha} \mathcal{R}^{\alpha i} u^-_i +
		\\&
		+16 (\bar{\theta}^+\bar{\theta}^-) \theta^{-\beta} \bar{\psi}^i_\beta u^-_i
		+
		8 (\bar{\theta}^-)^2 \theta^{+\beta} \bar{\psi}^i_\beta u^-_i
		-
		8 (\bar{\theta}^-)^2 \theta^{-\beta} \bar{\psi}^i_\beta u^+_i
		\\&
		-
		8i (\bar{\theta}^-)^2 (\theta^+\theta^-) \bar{\theta}^{+}_{\dot{\alpha}} \bar{\mathcal{R}}^{\dot{\alpha}i} u^-_i
		-
		4i (\bar{\theta}^+)^2 (\theta^-)^2 \bar{\theta}^{-}_{\dot{\alpha}} (\bar{\mathcal{R}}^{\dot{\alpha}i} + 2 \partial^{\dot{\alpha}\beta}\bar{\psi}_\beta^i) u^-_i
		\\
		& -4i (\theta^-)^4 \bar{\theta}^{+}_{\dot{\alpha}}
		\bar{\mathcal{R}}^{\dot{\alpha}i} u^+_i
		+
		8 (\theta^-)^4 (\bar{\theta}^+)^2 \theta^{+\rho}
		\partial_{\rho\dot{\alpha}} \bar{\mathcal{R}}^{\dot{\alpha}i} u^-_i.
	\end{split}
\end{equation}
\end{subequations}

\textit{\textbf{Auxiliary fermionic field $\rho_\beta^i$:}}
\begin{subequations}
\begin{equation}
	G_{(\rho)}^{++5} = 8 (\bar{\theta}^+)^2 \theta^{+\beta} \rho^i_\beta u^-_i
	-
	8
	(\theta^+)^2 \bar{\theta}^{+\dot{\beta}} \bar{\rho}^i_{\dot{\beta}} u^-_i,
\end{equation}
\begin{equation}
	\begin{split}
		G^{--5}_{(\rho)} = &\quad
		8(\bar{\theta}^+\bar{\theta}^-) \theta^{-\beta} \rho^i_\beta u^-_i
		+
		4 (\bar{\theta}^-)^2 \theta^{+\beta} \rho_\beta^i u^-_i
		-
		4 (\bar{\theta}^-)^2 \theta^{-\beta}\rho_\beta^i u^+_i
		\\&
		-8i (\bar{\theta}^-)^2 (\theta^+ \theta^-) \bar{\theta}^{+\dot{\sigma}} \partial_{\dot{\sigma}}^\beta \rho_\beta^i u^-_i
		+
		8i (\bar{\theta}^+)^2 (\theta^-)^2 \bar{\theta}^{-\dot{\sigma}} \partial_{\dot{\sigma}}^\beta \rho_\beta^i u^-_i
		\\&
		-16i (\bar{\theta}^-)^2 \theta^{-(\beta} \theta^{+\sigma)} \bar{\theta}^{+\dot{\sigma}} \partial_{\sigma\dot{\sigma}} \rho_{\beta}^i u^-_i
		\\& +
		8i (\theta^-)^4 \bar{\theta}^{+\dot{\sigma}} \partial^\beta_{\dot{\sigma}} \rho_\beta^i u^+_i
		- 4(\theta^-)^4 (\bar{\theta}^+)^2 \theta^{+\beta} \Box \rho_\beta^i u^-_i
		\\&
		-
		8 (\theta^+\theta^-) \bar{\theta}^{-\dot{\beta}} \bar{\rho}^i_{\dot{\beta}} u^-_i
		-
		4 (\theta^-)^2 \bar{\theta}^{+\dot{\beta}} \bar{\rho}^i_{\dot{\beta}} u^-_i
		+
		4 (\theta^-)^2 \bar{\theta}^{-\dot{\beta}} \bar{\rho}^i_{\dot{\beta}} u^+_i
		\\& -8i (\theta^-)^2(\bar{\theta}^+ \bar{\theta}^-) \theta^{+\sigma} \partial_\sigma^{\dot{\beta}} \bar{\rho}_{\dot{\beta}}^i u^-_i
		+
		8i (\theta^+)^2 (\bar{\theta}^-)^2 \theta^{-\sigma} \partial_\sigma^{\dot{\beta}} \bar{\rho}^i_{\dot{\beta}} u^-_i
		\\&
		+
		16 i (\theta^-)^2 \bar{\theta}^{-(\dot{\beta}} \bar{\theta}^{+\dot{\sigma})} \theta^{+\sigma} \partial_{\sigma\dot{\sigma}} \bar{\rho}^i_{\dot{\beta}} u^-_i
		\\&
		+8i (\theta^-)^4 \theta^{+\sigma} \partial_\sigma^{\dot{\beta}} \bar{\rho}_{\dot{\beta}}^i u^+_i
		+4 (\theta^-)^4 (\theta^+)^2 \bar{\theta}^{+\dot{\beta}} \Box \bar{\rho}^i_{\dot{\beta}}u^-_i.
	\end{split}
\end{equation}
\end{subequations}
\textit{\textbf{Auxiliary fermionic field $\chi_\beta^i$:}}
\begin{subequations}
\begin{equation}
	G^{++\alpha\dot{\alpha}}_{(\chi)} =
	- 4i (\theta^+)^4 \bar{\theta}^{-\dot{\alpha}} \chi^{\alpha i} u^-_i
	+
	4i (\theta^+)^4 \theta^{-\alpha} \bar{\chi}^{\dot{\alpha}i}u^-_i,
\end{equation}
\begin{equation}
	\begin{split}
		G^{--\alpha\dot{\alpha}}_{(\chi)} =&\quad\;
		4i (\bar{\theta}^-)^2 (\theta^+\theta^-) \bar{\theta}^{+\dot{\alpha}} \chi^{\alpha i}u^-_i
		-
		2i (\theta^-)^2 (\bar{\theta}^+)^2 \bar{\theta}^{-\dot{\alpha}} \chi^{\alpha i} u^-_i
		\\&
		-2i (\theta^-)^4 \bar{\theta}^{+\dot{\alpha}} \chi^{\alpha i} u^+_i
		-
		4 (\theta^-)^4 (\bar{\theta}^+)^2 \theta^{+\rho} \partial_\rho^{\dot{\alpha}} \chi^{\alpha i}u^-_i
		\\
		& -4i (\theta^-)^2 (\bar{\theta}^+ \bar{\theta}^-) \theta^{+\alpha} \bar{\chi}^{\dot{\alpha}i} u^-_i
		+
		2i (\bar{\theta}^-)^2 (\theta^+)^2 \theta^{-\alpha} \bar{\chi}^{\dot{\alpha}i}u^-_i
		\\
		&
		+
		2i (\theta^-)^4 \theta^{+\alpha} \bar{\chi}^{\dot{\alpha}i}u^+_i
		-
		4 (\theta^-)^4 (\theta^+)^2 \bar{\theta}^{+\dot{\rho}} \partial_{\dot{\rho}}^\alpha \bar{\chi}^{\dot{\alpha}i} u^-_i,
	\end{split}
\end{equation}

\medskip

\begin{equation}
	G^{++5}_{(\chi)} = -2i (\theta^+)^4 \theta^{-\alpha}\chi_\alpha^i u^-_i
	+
	2i (\theta^+)^4 \bar{\theta}^{-\dot{\alpha}}\bar{\chi}_{\dot{\alpha}}^i u^-_i,
\end{equation}
\begin{equation}
	\begin{split}
		G^{--5}_{(\chi)}
		=&\quad
		2i (\theta^-)^2 (\bar{\theta}^+ \bar{\theta}^-)  \theta^{+\alpha}\chi^i_\alpha u^-_i
		-
		i (\theta^+)^2 (\bar{\theta}^-)^2 \theta^{-\alpha} \chi^i_\alpha u^-_i
		\\&- i (\theta^-)^4 \theta^{+\alpha}\chi^i_\alpha u^+_i
		-
		2 (\theta^-)^4 (\theta^+)^2 \bar{\theta}^{+\dot{\rho}} \partial^\alpha_{\dot{\rho}} \chi^i_\alpha u^-_i +
		\\&
		-2i (\bar{\theta}^-)^2 (\theta^+\theta^-) \bar{\theta}^{+\dot{\alpha}} \bar{\chi}^i_{\dot{\alpha}}u^-_i
		+
		i (\bar{\theta}^+)^2 (\theta^-)^2 \theta^{-\dot{\alpha}} \bar{\chi}^i_{\dot{\alpha}} u^-_i
		\\
		& +i (\theta^-)^4 \bar{\theta}^{+\dot{\alpha}} \bar{\chi}^i_{\dot{\alpha}} u^+_i
		-
		2 (\theta^-)^4 (\bar{\theta}^+)^2 \theta^{+\rho} \partial_\rho^{\dot{\alpha}} \bar{\chi}^i_{\dot{\alpha}} u^-_i.
	\end{split}
\end{equation}
\end{subequations}


\end{document}